\documentclass[usenatbib]{mnras}

\usepackage{newtxtext,newtxmath}
\usepackage[T1]{fontenc}
\usepackage{ae,aecompl}

\usepackage[normalem]{ulem}
\usepackage{graphicx}	% Including figure files
\usepackage{amsmath}	% Advanced maths commands
\usepackage{pdflscape}
\usepackage{float}
\usepackage{epstopdf}
\usepackage{hyperref}
\usepackage{booktabs}
\title[T2Cs and ACs in K2]{Type II and anomalous Cepheids in the \textit{Kepler K2} mission}

\author[M. I. Jurkovic et al.]
{
Monika I. Jurkovic$^{1}$\thanks{E-mail: mojur@aob.rs},
Emese Plachy$^{2,3,4, 5}$\thanks{E-mail: plachy.emese@csfk.org}, 
L\'{a}szl\'{o} Moln\'{a}r$^{2,3,4, 5}$, \newauthor
Martin A. T. Groenewegen$^{6}$, Attila B\'{o}di$^{2,3,4}$,
Pawel Moskalik$^{7}$ 
and R\'{o}bert Szab\'{o}$^{2,3,4,5}$
\\
$^{1}$Astronomical Observatory of Belgrade, Volgina 7., 11060 Belgrade, Serbia\\
$^{2}$Konkoly Observatory, Research Centre for Astronomy and Earth Sciences, E\"otv\"os Lor\'and Research Network (ELKH), \\
Konkoly Thege Mikl\'os \'ut 15-17, H-1121 Budapest, Hungary\\
$^{3}$CSFK, MTA Centre of Excellence, Konkoly Thege Mikl\'os \'ut 15-17, H-1121 Budapest, Hungary\\
$^{4}$MTA CSFK Lend\"ulet Near-Field Cosmology Research Group\\
$^{5}$ELTE E\"otv\"os Lor\'and University, Institute of Physics, P\'azm\'any P\'eter s\'et\'any 1/A, H-1117 Budapest, Hungary\\
$^{6}$Royal Observatory of Belgium, Ringlaan / Avenue Circulaire 3, 1180 Brussels, Belgium\\
$^{7}$Nicolaus Copernicus Astronomical Centre of the Polish Academy of Sciences, ul. Bartycka 18, 00-716 Warszawa, Poland\\
}

\date{Accepted XXX. Received YYY; in original form ZZZ}

\pubyear{2022}

\hypersetup{draft}
\begin{document}
\label{firstpage}
\pagerange{\pageref{firstpage}--\pageref{lastpage}}
\maketitle

\begin{abstract}
We present the results of the analysis of Type II and anomalous Cepheids using the
data from the \textit{Kepler K2} mission. The precise light
curves of these pulsating variable stars are the key to study the details of their pulsation, such as the period-doubling effect or the presence of additional modes. We applied
the Automated Extended Aperture Photometry (autoEAP) to obtain the light curves
of the targeted variable stars which were observed. The light curves
were Fourier analyzed. We investigated twelve stars observed by the K2 mission, seven
Type II and five anomalous Cepheids. Among the Type II Cepheids EPIC 210622262
shows period-doubling, and four stars have modulation present in their light curves
which are different from the period-doubling effect. We calculated the high-order Fourier parameters for the short-period Cepheids. We
also determined physical parameters by fitting model atmospheres to the spectral energy distributions. The determined distances using the parallaxes measured by the
Gaia space telescope have limited precision below 16 mag for these types of pulsating
stars, regardless if the inverse method is used or the statistical method to calculate the distances. The BaSTI evolutionary models
were compared to the luminosities and effective temperatures. Most of the Type II
Cepheids are modeled with low metallicity models, but for a few of them solar-like
metallicity ([Fe/H]=0.06) model is required. The anomalous Cepheids are compared
to low-metallicity single stellar models. We do not see signs of binarity among our
sample stars.
\end{abstract}


%%%%%%%%%%%%%%%%% BODY OF PAPER %%%%%%%%%%%%%%%%%%

\section{Introduction}
\label{sec:intorduction}

Type II Cepheids (T2Cs) are low mass (~0.5 M$_{\odot}$), radially pulsating stars. They form three subgroups, which can be separated by their periods: BL Herculis (BLH, with periods 1<P<5 days), W Virginis, including peculiar W Virginis stars ((p)WVir), with periods 5<P<20 days, and the RV Tauri (RVT, with periods greater than 20 days). The limits in between the periods follow the OGLE (Optical Gravitational Lensing Experiment\footnote{\url{http://ogle.astrouw.edu.pl/}})
classification scheme, e.g. see catalog of the OGLE-IV variables toward the Galactic Centre \citep{2017_OGLE-IV_Bulge_variables}. 

Anomalous Cepheids (ACs) are variable stars with masses up to 1.3 M$_{\odot}$ \citep{Groenewegen_2017b}, which can  pulsate in the fundamental or first radial overtone modes. Their pulsation periods range from 0.3 to 2.4 days \citep{OGLE_LMC_T2C_ACEP_2008, OGLEIV_ACEP_2015}. Their origin can have two different scenarios - they are either a result of a metal-poor single stellar evolution or they are the end result of the binary evolution \citet{Fiorentino_2012}.

Both of these variable types occupy the instability strip (IS) of the Hertzsprung-Russell diagram (HRD) \citep{Catelan_book_2015}, and they overlap in the short period range, but they form separate period-luminosity relations (PL-relations), see for example \citet{2022ApJ...927...89W, 2021ASPC..529..259B, 2021MNRAS.501..875D, 2020JApA...41...23B, 2019A&A...625A..14R, Groenewegen_2017b, 2017AJ....153..154B, OGLEIV_ACEP_2015, 2011MNRAS.413..223M, OGLE_LMC_T2C_ACEP_2008}, and references within. In the article by \citet{2020A&A...644A..95B} it is stated that the PL-relations of T2Cs, together with the RR Lyrae (RRL, short period pulsating stars) variable stars, can also be used as extragalactic distance indicators as are classical Cepheids used. The old stellar populations, such as T2Cs, probe the galaxies in different stages of interstellar material enrichment and star formation, thus providing an insight into the timeline of the galactic evolution. ACs are mostly found in dwarf galaxies, and could be used as independent distance indicators as well. From the evolutionary point of view, these are two very different types of stars, arriving onto the IS in different ways. T2Cs are in a late stage of their evolution, as they have reached the horizontal branch (HB) and they continue their last crossing of the IS before they loose all their outer shell and become a white dwarf (see \citet{2020A&A...644A..95B} for details). ACs are core He-burning stars that have higher masses than T2Cs, so they are brighter,  \citep[see, e.g.,][]{2006A&A...460..155F}.

The theoretical hydrodynamic models of \citet{Smolec_2012,  Smolec_2014, Smolec_2016, Buchler_1992, Moskalik_1993, Moskalik_Buchler_1990, Kovacs_Buchler_1988} show that BLHs and WVirs exhibit non-linear phenomena such as period-doubling (PD), but long-term, precise observations are required to detect these. Clear detection of PD in a BLH star was reported, for the first time, by \citet{OGLEIII_T2C_bulge_2011} and \citet{Smolec_2012_PD_detect} in the case of OGLE-BLG-T2CEP-279.

Cycle-to-cycle changes in WVir stars were already known (see \citet{Percy_book}), but one must be careful about interpreting the observed phenomena. In the case of W Virginis itself \citet{Henden_1980} published their results stating that the star shows cycle-to-cycle variation, but the reanalysis of \citet{Plachy_2017} confirmed that the observed phenomenon is PD. 

The analysis of 924 T2Cs in the Bulge of the Milky Way from OGLE-IV data sets was published by \citet{Smolec_2018}. The detailed and in-depth insight into the 
nature of the pulsation of these stars showed that T2Cs can exhibit radial double-mode pulsation (pulsating simultaneously in the fundamental and first overtone modes). Then \citet{2019ApJ...873...43S} found two candidates for T2Cs pulsating in the first overtone. The phenomenon of period doubling appears in WVir stars with periods longer than 15 days and it becomes common and sometimes irregular in the RVT regime, but it is less common among BLH stars. Periodic modulation of the pulsation has been detected in all subgroups of T2Cs. 

The space-based photometry of T2Cs was very limited before the \textit{K2} mission of the \emph{Kepler} space telescope \citep{2010Sci...327..977B,2016RPPh...79c6901B}. The Convection, Rotation and planetary Transits mission \citep[\emph{CoRoT},][]{2007AIPC..895..201B}, operational between 2006 and 2013, observed only one T2C, CoRoT 0659466327, in which \citet{Poretti_2015} found no additional frequencies beside the fundamental period and its harmonics. The original \emph{Kepler} field-of-view contained one T2C, the RVT star DF Cygni. Detailed analyses of its behavior were published by \citet{Bodi_2016, Vega_2017, Plachy_2018, Manick_2019}, finding additional, long-term period change due to the presence of a binary companion and a circumstellar disk, as well as the presence of chaos in its pulsation. \citet{Manick_2019} found that DF Cyg is most probably a post red giant branch (post-RGB) binary. \citet{Plachy_2017} published the first results of two T2Cs from the \textit{K2} mission, namely KT Sco and M80 V1, where they found PD in the latter star.

In the case of ACs \citep{Stetson_LeoI_2014} already described a possible Blazhko modulated (longer period variation of the amplitude in the minimum and maximum position of the light curve resulting in a light curve shape change as well observed in RRL type variables) or multiple-mode anomalous Cepheid, V129, in the dwarf spheroidal galaxy Leo I. \citet{2018SerAJ.197...13J} found Blazhko-like modulation in FY Virginis, a Milky Way AC. \citet{2019ApJS..244...32P} found four anomalous Cepheids among the RRL candidate stars in the \textit{K2} sample. Among these stars two show signs of a Blazhko-effect, making this the first definite discovery of this phenomenon in anomalous Cepheids. \citet{2020ApJ...901L..25S} found the first AC, OGLE-GAL-ACEP-091, pulsating in multiple modes and \citet{2021ApJS..253...11P} showed that the AC star XZ Cet is not just pulsating in a first-overtone but also has a strong secondary mode. 

In this paper we use data from the \emph{Kepler} space telescope during its extended \textit{K2} mission \citep{2014PASP..126..398H}. We examine which of the above mentioned phenomena can be detected in the T2Cs and ACs in the observed \textit{K2} Campaigns. Section~\ref{sec:data} describes the data selection process. In Subsection~\ref{subsec:targets}, we give the description of the target selection, in  Subsection~\ref{subsec:photometry}, we describe how the photometry was performed. The Fourier analysis of the sampled stars is described in detail in Subsection~\ref{subsec:Fourier_analysis}, and the Fourier analysis is explained. In Subsubsections~\ref{subsec:ACs}, \ref{subsec:BLHs} and \ref{subsec:WVirs} results of the Fourier analysis of individual stars is presented. In Section~\ref{sec:physical_param}, we use the Spectral Energy Distribution method to derive the physical parameters of each star, and then use it to construct the Hertzsprung--Russell Diagram (see Subsection~\ref{subsec:hrd}), the Color--Magnitude Diagram (see Subsection~\ref{subsec:CMD}) and the Period--Radius relation (see Subsection~\ref{subsec:PR_relation}). Section~\ref{sec:PL_relation} deals with the distance determination issue. In particular in Subsection~\ref{subsec:PL_T2C_AC}, we show the Period-Magnitude relations, while in Subsection~\ref{subsec:plx_vs_d}, we discuss how distances derived from the 
\textit{Gaia} space telescope should be treated with caution. Finally, Section~\ref{sec:conclusions} contains the summary of our results.

%%%%%%%%%%%%%%%%%%%%%%%%%%%%%%%%%%%%%%%%%%%%%%%%%%%%%%%%%%%%%%

\section{The \textit{Kepler - K2} dataset}
\label{sec:data}

%%%%%%%%%%%%%%%%%%%%%%%%%%%%%%%%%%%%%%%%%%%%%%%%%%%%%%%%%%%%%%%%

\subsection{Target selection}
\label{subsec:targets}

The \emph{Kepler} space telescope observed thousands of pre-selected targets during each campaign of the \textit{K2} mission\footnote{\url{https://archive.stsci.edu/missions-and-data/k2}, \url{https://keplergo.github.io/KeplerScienceWebsite/}} \citep{2014PASP..126..398H}, including several Cepheids. The Cepheid targets we proposed are based on the available classification from the literature. Targets were selected from the General Catalogue of Variable Stars (GCVS\footnote{\url{http://www.sai.msu.su/gcvs/gcvs/}}), the Catalina Sky Survey (CSS\footnote{\url{https://catalina.lpl.arizona.edu/}}) and the All Sky Automated Survey (ASAS\footnote{\url{http://www.astrouw.edu.pl/asas/?page=main}}) databases \citep{2017ARep...61...80S,Drake_2014,Pojmanski_1997}. Among the selected and observed targets only two T2Cs have been analysed in detail so far \citep{Plachy_2017}. 

We reviewed the remaining Cepheid targets searching for T2Cs and ACs among them.
We excluded the Cepheids in high stellar density fields from our investigations, such as the Galactic Bulge, observed in Campaigns 9 and 11, and the IC\,1613 dwarf galaxy observed in Campaign 8, as these stellar fields require dedicated photometric solutions that are beyond the scope of this paper. 

We found seven T2Cs and five ACs in the observed Campaigns, as shown in Table~\ref{tab:stars_in_K2}, which we analysed in detail. The table shows the number of the campaign in which the star was observed, the \emph{Kepler} identification number (EPIC ID), the positions of the stars (right ascension, RA, and declination, DEC), the pulsational period (P) in days that were available in the The International Variable Star Index (VSX\footnote{\url{https://www.aavso.org/vsx/}}), the average \textit{Kepler} magnitude of the light curve, the type of the variability determined in this work, and other known names from catalogs for the observed stars. Two long period T2C candidates, belonging to the RVT subgroup (see Table~\ref{tab:other_T2C}), were excluded from this study, since the length of the observed \textit{K2} data was insufficient for further investigation. We revealed that several former Cepheid candidates were wrongly classified, and they are listed in Table~\ref{tab:other_stars}.

\begin{table*}
	\centering
	\caption{The T2Cs and ACs examined in \textit{K2} data. The first column gives the Campaign number in which the star was observed in the \textit{K2} mission, while its EPIC catalog number is in the second column. The positions of the stars on the sky is given in the third and fourth columns (right ascension, RA, and declination, DEC) from the $^1$\citep{2020yCat.1350....0G}. The fifth column is the fundamental pulsation period (P) in days of the variable stars from the VSX database. The type of the variability - confirmed in this paper - is in the sixth column. The seventh column contains the other know names of the stars in the sample.}
	\label{tab:stars_in_K2}
	\begin{tabular}{l ccclcll}
		Camp. & EPIC & RA$^1$ & Dec$^1$ & P$_{VSX}$ & <Kp> & Type in & Other name \\
		 & ID & [h m s] & [$^\circ$ ' "] & [days] & [mag] & this article & \\
		\hline
        2 & 202862302 & 16 36 52.85 & -28 05 34.21 & 1.956 & 12.926 & AC & V1287 Sco\\
        \hline
        4 & 210622262 & 04 20 01.79 & +17 16 45.83 & 16.635 & 16.882 & WVir & CSS\_J042001.7+171645\\
        \hline
        7 & 215881928 & 18 59 37.23 & -23 21 52.28 & 1.835 & 14.606 & BLH & V839 Sgr\\
        7 & 217235287 & 19 16 10.99 & -20 55 55.82 & 1.259 & 15.155 & BLH & V527 Sgr\\
        7 & 217693968 & 18 48 09.79 & -20 07 35.61 & 16.206 & 13.289 & WVir & V377 Sgr\\
        7 & 217987553 & 19 06 26.94 & -19 36 35.28 & 13.429 & 12.482 & WVir  & V1077 Sgr\\
        7 & 218128117 & 19 34 34.67 & -19 21 39.99 & 2.119 & 12.735 & AC & ASAS J193435-1921.7 \\ 
        7 & 218642654 & 19 06 03.13 & -18 25 41.65 & 13.775 & 12.166 & WVir  & V410 Sgr\\
        \hline
        12 & 246015642 & 23 39 54.14 & -09 05 01.74 & 1.071 & 15.399 & AC & CSS\_J233954.1-090502\\
        12 & 246333644 & 23 22 33.10 & -02 23 40.08 & 1.287 & 17.792 & AC & CSS\_J232233.0-022339\\
        12 & 246385425 & 23 15 26.53 & -01 22 28.66 & 1.502 & 17.972 & AC & CSS\_J231526.5-012228\\
        \hline
        13 & 247445057 & 05 05 14.25 & +21 45 48.91 & 13.943 & 12.355 & WVir & VZ Tau\\
        \hline
       \end{tabular}
\end{table*}

%%%%%%%%%%%%%%%%%%%%%%%%%%%%%%%%%%%%%%%%%%%%%%%%%%%%%%

\subsection{Photometry}
\label{subsec:photometry}

\begin{figure}
    \centering
    \includegraphics[width=8.4cm]{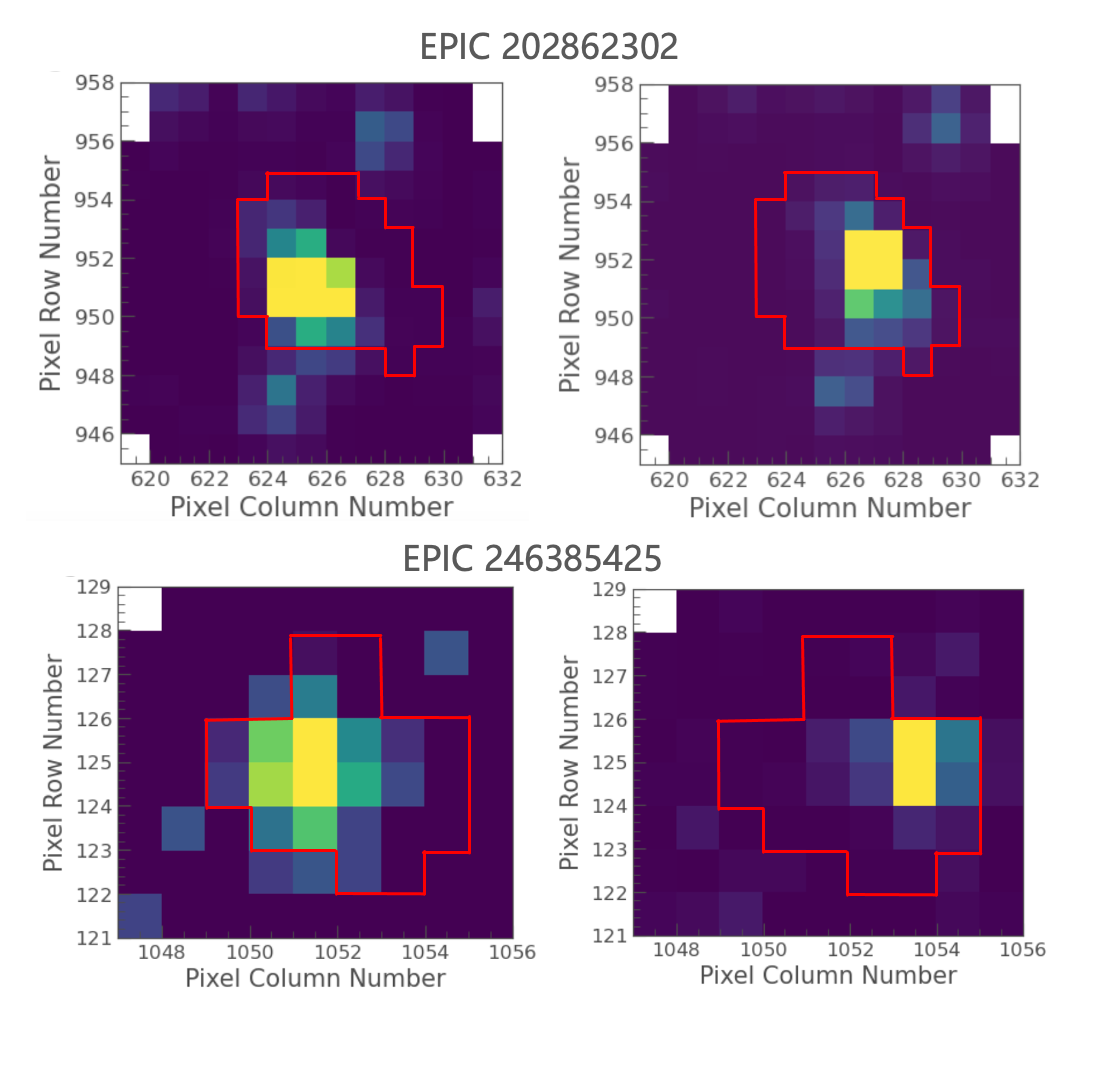}
    \caption{Examples of Target Pixel Frame (TPF) images for the stars EPIC 202862302 and EPIC 246385425 in extreme positions. The aperture pixel masks highlighted in red were determined by the autoEAP pipeline.}
    \label{fig:pixel_mask}
\end{figure}

The \textit{K2} mission had continued to deliver photometric data, but with lower quality compared to the original \emph{Kepler} mission (after losing two reaction wheels that controlled the stability of the spacecraft). Systematic noise and instrumental signals resulted in light curves that needed to be cleaned from these artefacts. Additionally, the pointing deficiencies of the space craft created characteristic patterns in the data throughout each Campaign. All these additional signals could mimic real dynamical features of pulsating stars, such as amplitude modulation and cycle-to-cycle changes. Several pipelines have been developed to lessen or eliminate the instrumental issues. One of these has been optimized to separate the RR Lyrae type light variation with the highest possible quality, called the Extended Aperture Photometry (EAP, \citealt{2019ApJS..244...32P}). In the EAP the algorithm chooses the pixels of the aperture by searching for the light variability of the target star at any time during the observation in them. That results in an extended aperture that covers the star at any position as it is demonstrated in Figure~\ref{fig:pixel_mask}, where we display the pixel mask or a Target Pixel Frame (TPF) of the example stars at the extreme positions. Here we used the newest, automated version of EAP, the autoEAP pipeline \citep{autoeap, 2021arXiv211207496B}. The \textit{K2} Systematic Correction (K2SC, \citet{k2sc}) was applied to further improve the quality of our light curves.

%%%%%%%%%%%%%%%%%%%%%%%%%%%%%%%%%%%%%%%%%%%

\subsection{Analysis using Fourier decomposition}
\label{subsec:Fourier_analysis}

\begin{table*}
\caption{The calculated Fourier parameters of the stars from the \textit{K2} sample.}
\label{tab:fou}
\begin{tabular}{|p{1.8cm}|p{1.3cm}|p{1.3cm}|p{1.3cm}|p{1.3cm}|p{1.3cm}|p{1.3cm}|p{1.3cm}|p{1.3cm}|}
\hline
EPIC ID & $f_1$ &$\sigma f_1$ &P$_1$& $\sigma$P$_1$ &A$_1$ & $\sigma$A$_1$ &  Harmonics&\\
\hline
202862302	&	0.5113902	&	0.0000017	& 1.9554540	&	0.0000065 &	0.16840	&	0.00004	& 17	&		\\
210622262	&	0.0599405	&	0.0000125	& 16.6832109	&	0.0034791	&	0.48477	&	0.00073	& 4	&		\\
215881928	&	0.5448335	&	0.0000012	& 1.8354231	&	0.0000040	&	0.28267	&	0.00007	& 20	&		\\
217235287	&	0.7943928	&	0.0000010	& 1.2588231	&	0.0000016	&	0.33988	&	0.00005	& 30	&		\\
217693968	&	0.0615492	&	0.0000053	& 16.2471649	&	0.0013990	&	0.40777	&	0.00028	& 3	&		\\
217987553	&	0.0744979	&	0.0000059	& 13.4231972	&	0.0010631	&	0.41338	&	0.00038	& 3	&		\\
218128117	&	0.4718607	&	0.0000012	& 2.1192695	&	0.0000054	&	0.32199	&	0.00007	& 48	&		\\
218642654	&	0.0725999	&	0.0000033	& 13.7741237	&	0.0006261	&	0.37290	&	0.00021	& 3	&		\\
246015642	&	0.9337528	&	0.0000010	& 1.0709473	&	0.0000011	&	0.23580	&	0.00005	& 25&		\\
246333644	&	0.7769263	&	0.0000020	& 1.2871234	&	0.0000033	&	0.39774	&	0.00028	& 25&		\\
246385425	&	0.6657856	&	0.0000026	& 1.5019850	&	0.0000059	&	0.35242	&	0.00030	&30	&		\\
247445057	&	0.0715535	&	0.0000054	& 13.9755568	&	0.0010547	&	0.45317	&	0.00044	& 3	&		\\
\hline
EPIC ID & R$_{21}$ & $\sigma$R$_{21}$ & $\varphi_{21}$ & $\sigma\varphi_{21}$  & R$_{31}$ & $\sigma$R$_{31}$ & $\varphi_{31}$ &$\sigma\varphi_{31}$ \\
\hline
202862302	&	0.3497	&	0.0002	&	4.5784	&	0.0632	&	0.1119	&	0.0003	&	2.8506	&	0.0951		\\
210622262	&	0.1567	&	0.0017	&	6.1032	&	0.5019	&	0.0862	&	0.0017	&	3.5409	&	0.7547		\\
215881928	&	0.2027	&	0.0003	&	5.1638	&	0.0539	&	0.123	&	0.0003	&	2.4219	&	0.0812		\\
217235287	&	0.2247	&	0.0002	&	4.5201	&	0.0439	&	0.1543	&	0.0002	&	3.6138	&	0.0658		\\
217693968	&	0.0574	&	0.0008	&	1.2767	&	0.2379	&	0.0819	&	0.0009	&	2.5225	&	0.3571		\\
217987553	&	0.3849	&	0.0011	&	5.5215	&	0.2654	&	0.1000	&	0.0010	&	2.3702	&	0.3971		\\
218128117	&	0.4513	&	0.0003	&	4.7151	&	0.0518	&	0.2095	&	0.0002	&	2.8334	&	0.0776		\\
218642654	&	0.2382	&	0.0006	&	6.1255	&	0.1492	&	0.0999	&	0.0006	&	2.5737	&	0.2230		\\
246015642	&	0.4450	&	0.0003	&	4.0950	&	0.0525	&	0.3264	&	0.0003	&	2.1435	&	0.0787		\\
246333644	&	0.5081	&	0.0007	&	4.1728	&	0.1061	&	0.3552	&	0.0007	&	2.1509	&	0.1589		\\
246385425	&	0.5264	&	0.0009	&	4.4309	&	0.1374	&	0.3355	&	0.0008	&	2.5007	&	0.2063		\\
247445057	&	0.1554	&	0.0011	&	0.5144	&	0.2902	&	0.1207	&	0.0009	&	2.2681	&	0.4331		\\
\hline
\end{tabular}
\end{table*}

We performed the Fourier analysis with our own \texttt{python} code using the Lomb-Scargle (LS) method \citep{Lomb,Scargle} to calculate the Fourier spectrum. We used the LS method as it is implemented in the \texttt{Astropy} package \citep{astropy:2013, astropy:2018}. The code first finds the frequency with the highest amplitude in the spectrum and after that does a consecutive pre-whitening with the integer multiples of that frequency. We used cosine-based Fourier series in the form of: 

\begin{equation}
    m (t) = m_0 + \sum_{i=1}{} A_i \cos(2\pi if t + \varphi_i),
	\label{eq:Fourier}
\end{equation}	
where $m (t)$ is the calculated decomposition model, $m_0$ is the average brightness in magnitude, $A_i$ is the $i$th semi-amplitude, $f$ is the frequency and $\varphi_i$ is the phase of the given component.

The pre-whitening was performed until the frequency of the $i$th harmonic was lower than the average Nyquist limit:
\begin{equation}
    f_{\rm Nyq} = \frac{1}{2<\Delta t>},
\end{equation}
where $<\Delta t>$ is the average duration between consecutive observations,
or the signal-to-noise ratio (S/N) of the given harmonic fell below 3 \citep{Breger93}. After the iteration stopped all the amplitudes and phases, along with the mean magnitude and frequency, were fine-tuned by fitting all the Fourier components simultaneously to the light curve. This optimization process started from the previously determined values.

The errors of the Fourier components were estimated with a bootstrap method as follows. We re-sampled the light curves a hundred times by randomly selecting 60\% of all measurements in each step. The free parameters were refitted to these subsets. The errors were calculated as the standard deviation of the distribution of the aggregated results of all the fits. 

The relative Fourier parameters, that are widely used for the description of the light curves, defined by \citep{Simon_1981, Simon_1986}, are the ratio of the amplitudes of the main frequency and its harmonics:
\begin{equation}
    R_{i1}=\frac{A_i}{A_1},
	\label{eq:R_i1}
\end{equation}

\noindent and the relative phase differences:
\begin{equation}
   \varphi_{i1}=\varphi_i - i\varphi_1,
	\label{eq:phi_ik}
\end{equation}
\noindent where $i=2, 3$.

\begin{figure*}
    \centering
    \includegraphics[width=\textwidth]{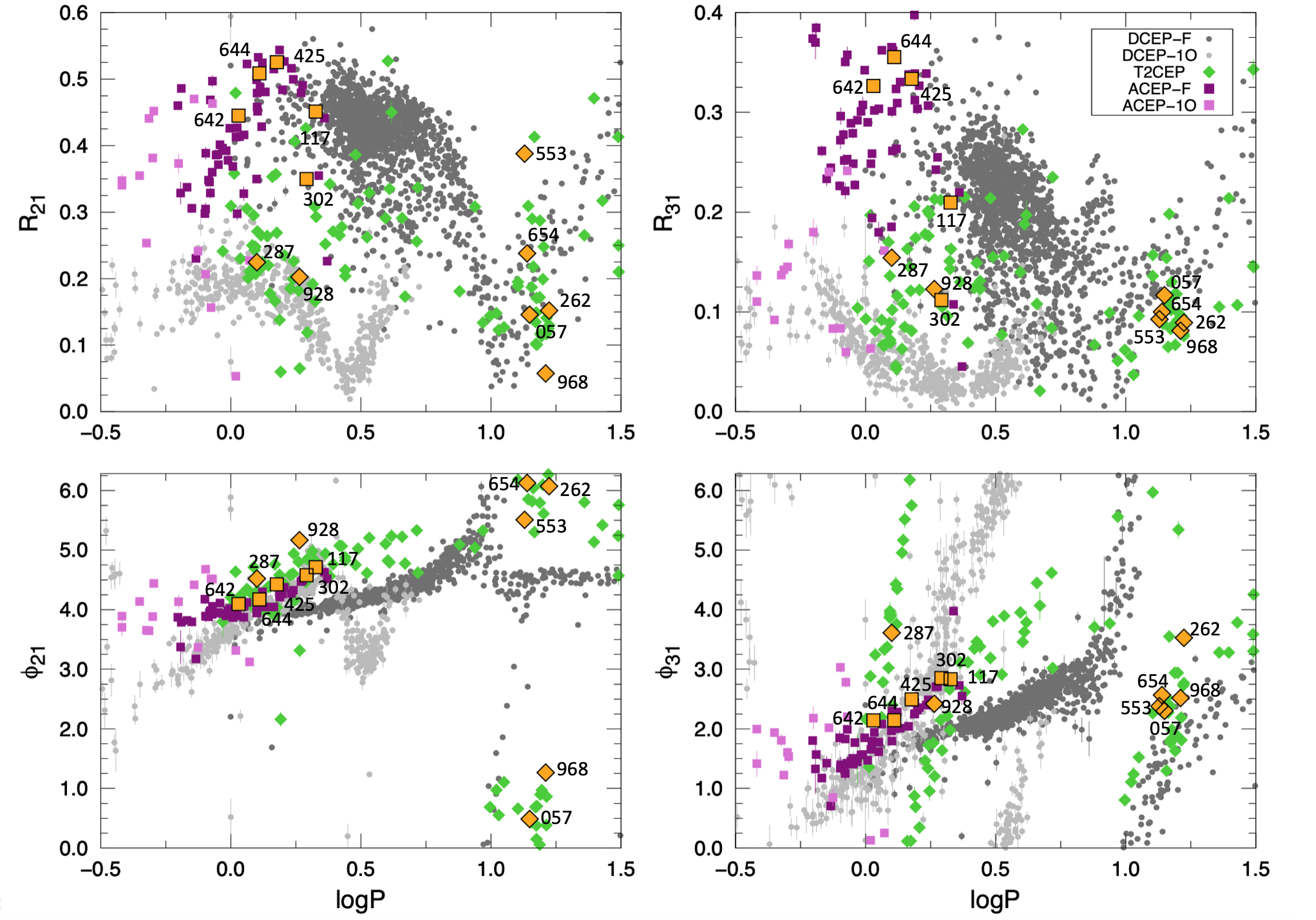}
    \caption{The cosine-based Fourier parameters ($R_{21}$, $R_{31}$, $\varphi_{21}$ and $\varphi_{31}$) versus the logarithm of the pulsational periods in days (P) of the \textit{K2} Cepheids (large orange symbols, and identified by the last three numbers of their EPIC identifier) plotted with the OGLE Large Magellanic Cloud Cepheid's Fourier parameters from the $V$ band data. Fundamental mode classical Cepheids (DCEP-F) are presented with dark \textbf{grey points}, first overtone classical Cepheids (DCEP-1O) are in light \textbf{grey points}, Type II Cepheids (T2C) are green diamonds, fundamental mode anomalous Cepheids (AC-F) are shown in dark purple squares and first overtone anomalous Cepheids (AC-1O) are presented with pink \textbf{squares}.}
    \label{fig:Fourier}
\end{figure*}

%%%%%%%%%%%%%%%%%%%%%%%%%%%%%%%%%%%%%%%%%%%%%%%%%%%%%%%

\subsection{Light curve features}
\label{subsec:LC_features}

\begin{figure*}
    \centering
    \includegraphics[width=\textwidth]{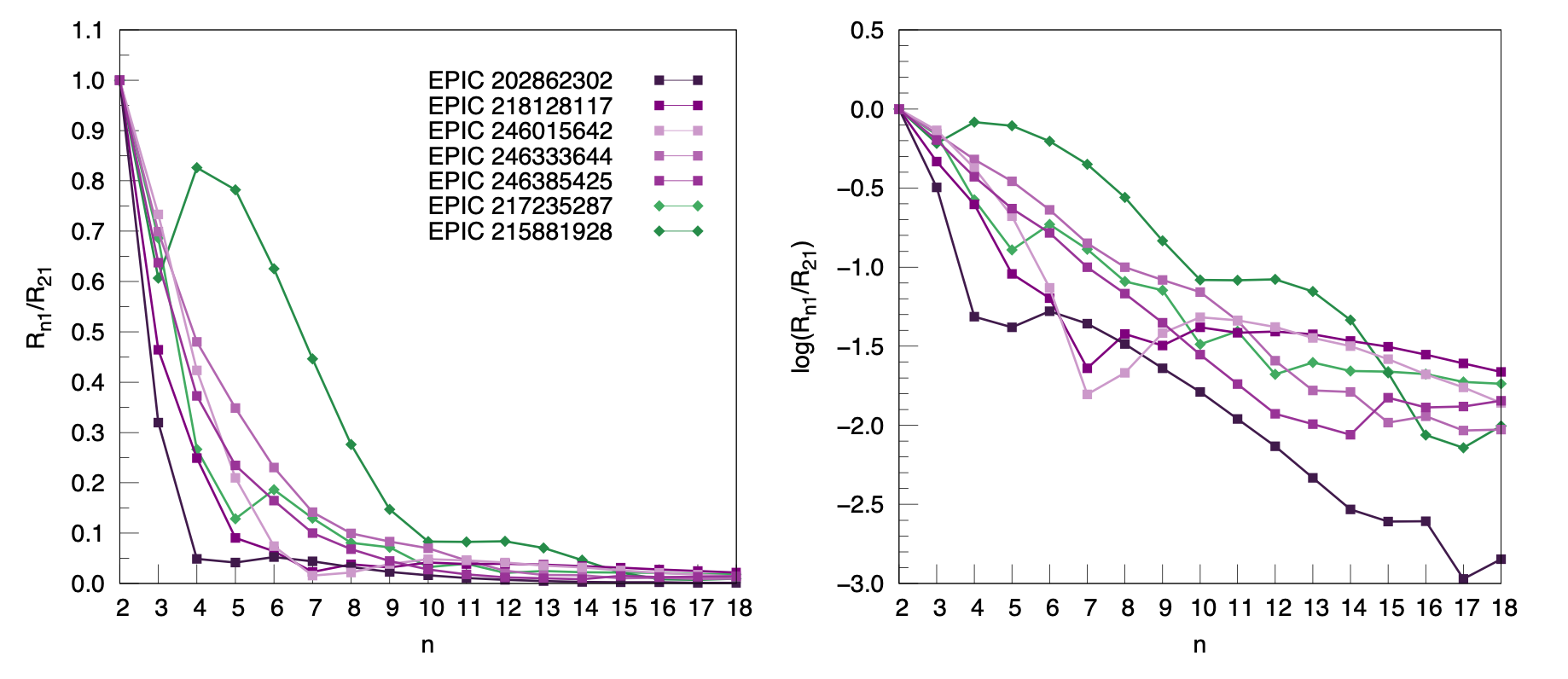}
    \caption{High-order Fourier relations of the amplitudes ($\frac{R_{n1}}{R_{21}}$) for the ACs (purple) and the BLH stars (green) in the \textit{K2} sample versus the number of detected harmonics. Values of the amplitudes ($\frac{R_{n1}}{R_{21}}$) are normalized and displayed on linear (left) and logarithmic (right) scales.}
    \label{fig:high}
\end{figure*}

\begin{figure*}
    \centering
    \includegraphics[width=15cm]{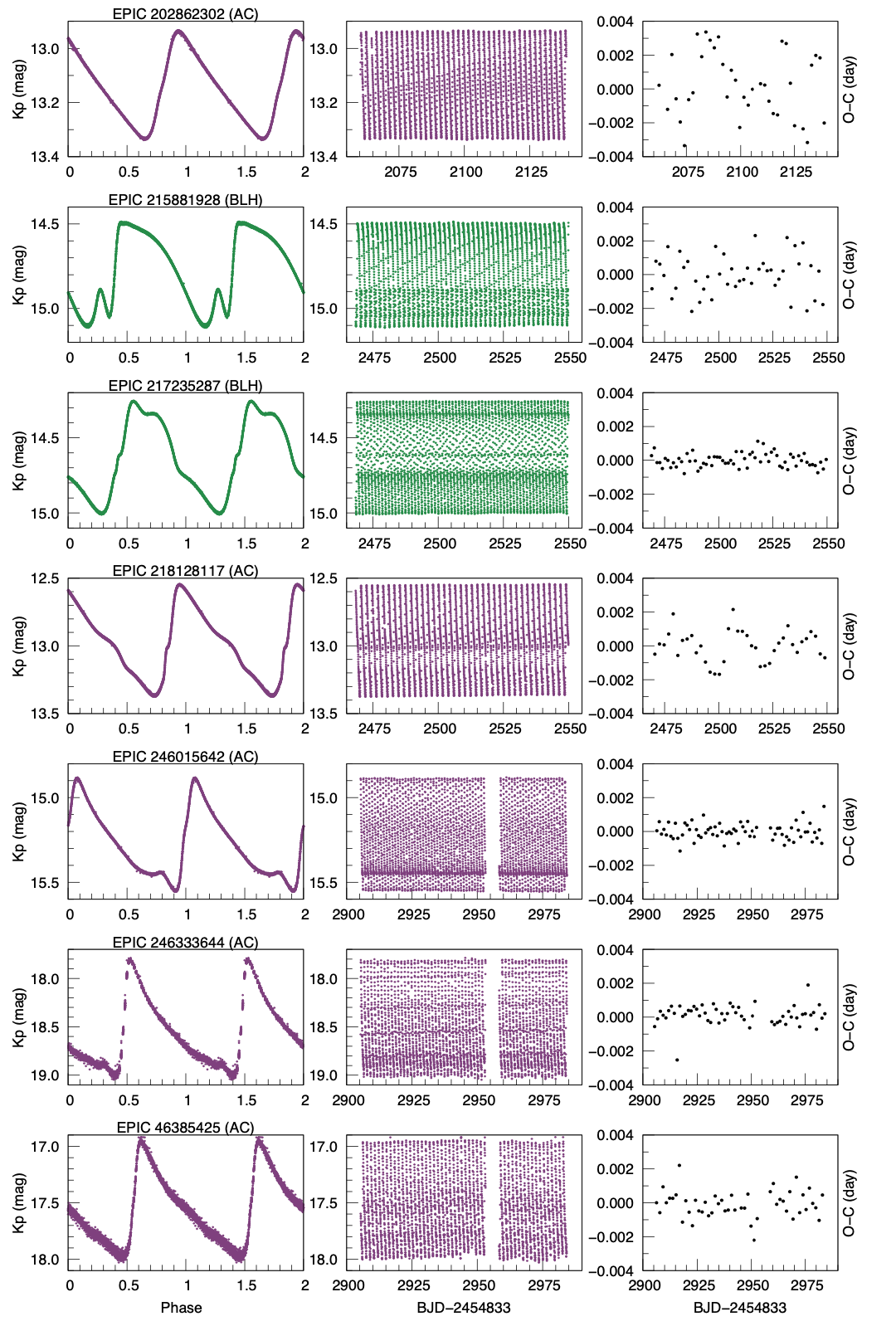}
    \caption{The phase folded light curves (left panel), observed light curves (middle panel) and O--C diagrams (right panel) of the ACs (purple) and BLHs (green) from the \textit{K2} data.}
    \label{fig:blh-ac}
\end{figure*}

\begin{figure*}
    \centering
    \includegraphics[width=0.9\textwidth]{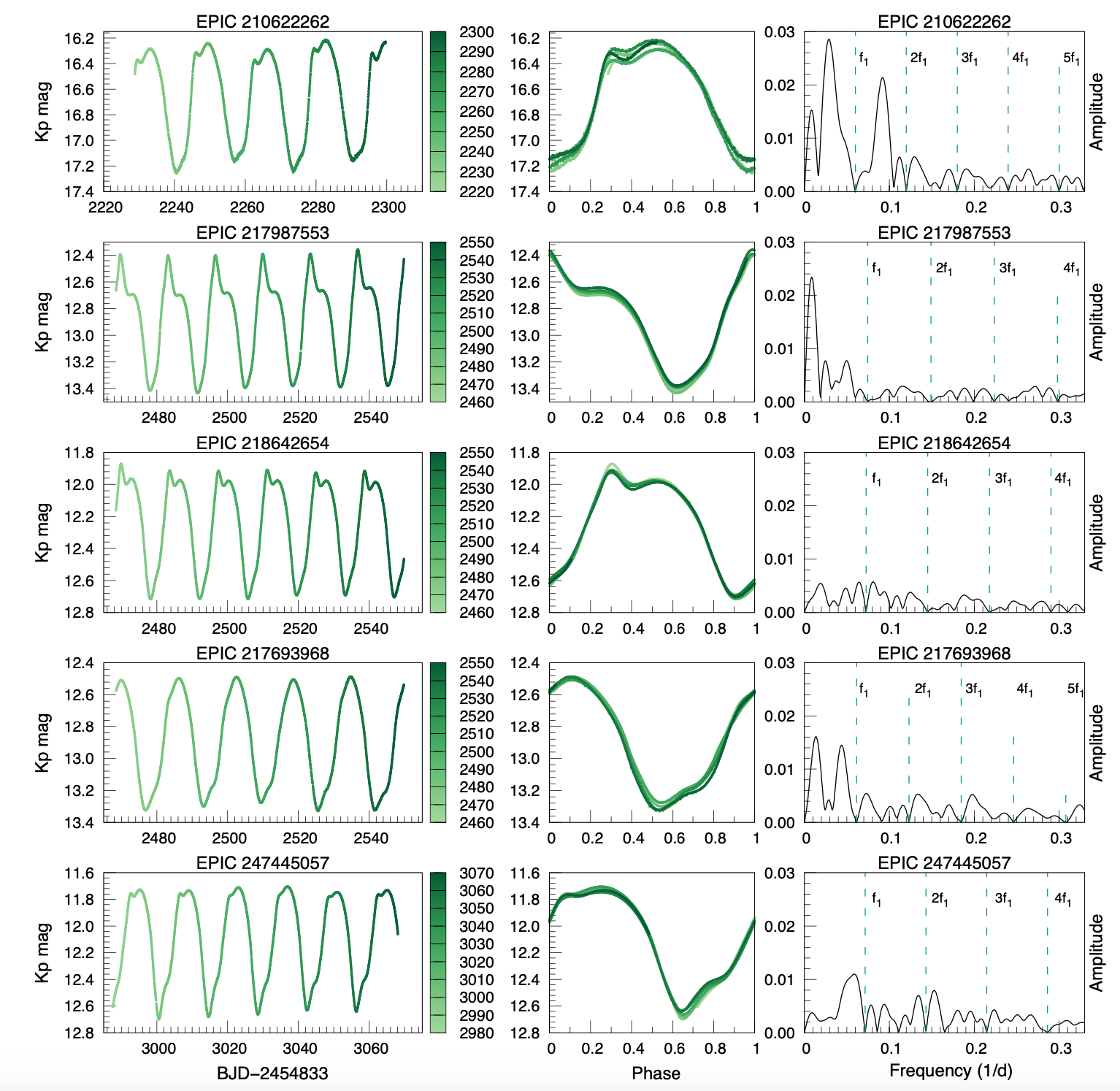}
    \caption{\textit{K2} light curves of WVir stars. On the left panel the total light curve is shown. The green scale goes from light green to dark green with time of the measurement (BJD-2454833) in days. The panel in the middle shows the phase folded light curves. The right hand panel shows the pre-whitened Fourier spectra (the frequency in [1/days] versus the amplitude in [mag]) of the light curves. The dashed green vertical lines show the position of the fundamental mode (f$_1$) and the consequent harmonics (2f$_1$, 3f$_1$, 4f$_1$ and 5f$_1$ in some cases).}
    \label{fig:wvir}
\end{figure*}

The Fourier parameters of the \textit{K2} stars are listed in Table \ref{tab:fou}. The number of harmonics were derived with the detection limit of S/N $\geq$ 3 and the Nyquist frequency (24.468 c/d).
These Fourier parameters were compared to those of the Cepheids measured by the OGLE Survey. We chose the Cepheid sample of the Large Magellanic Cloud (LMC), \citet{OGLE_LMC_T2C_ACEP_2008,2015AcA....65....1U}, where the clear Period-Luminosity relations validate the classification reliably.
At the time of writing this paper the OGLE database provides Fourier parameters only in $I$ photometric band. The $V$ band parameters, however, are more suitable for comparison, since they are close to the $K$ passband of the \textit{Kepler} space telescope. 
We downloaded the available $V$ band light curves and merged the data from OGLE-III and IV for the stars that were observed in both projects. Then we excluded those stars that have less than fifty measurements altogether.
For the rest, we calculated the Fourier coefficients with our \texttt{python} code mentioned above. The second harmonic peak turned out to be below our signal-to-noise threshold for a significant number of stars, which we then also excluded. For error estimation we chose a computationally less expensive method than the bootstrap. Errors were considered as square root of the variance from the diagonal of the covariance matrix, which was calculated by \texttt{scipy}'s \texttt{curve\_fit} method \citep{scipy}. The calculated $V$ band Fourier parameters are provided in the Appendix~\ref{app:OGLE_V_Fourier}.

In Figure~\ref{fig:Fourier}, we show the comparison of Fourier parameters for \textit{K2} and OGLE stars, where the parameters are displayed as a function of the logarithm of the period in days. Colors and symbols indicate the different variability types, and we can check whether the \textit{K2} stars are in the location where they should be according to their type. There are overlaps between the different groups, so the stars were checked individually in all four plots. Ideally, the position should clearly refer to a certain type at least on one of the plots. This is the case for the five WVir stars in the the $log P-\varphi_{21}$ plot, where they are clearly separated from the fundamental mode classical Cepheids (DCEP-F) with similar periods. The two BLH stars (marked as 287 and 928) are in the region of first overtone classical Cepheids (DCEP-1O) in the $log P-R_{21}$ plane, but are distant from them in at least two other panels. We can be also confident in the type of the three shorter period ACs, but the two longer ones, EPIC 218128117 and EPIC 202862302, are not distinct from BLH stars in any panels. This shows that Fourier parameters have limitations when we try to use them for classification purposes of the ACs and T2Cs. 

Investigating the Fourier parameters of higher orders, we may overcome the problem of precise classification. Therefore we calculated R$_{i1}$ up to the highest possible $i$ values for BLH and AC stars in our sample, and displayed the changes of these harmonic series in Figure \ref{fig:high} (both on linear and logarithmic scales). Unfortunately, no database for high-order Fourier parameters is available for comparison, but this can change with the massive analysis of Cepheids observed by the Transiting Exoplanet Survey (TESS) space telescope in the future. The first step has already been taken in the paper of \citet{2021ApJS..253...11P}, where nine ACs and one BLH stars were analysed. Clearly, the decrease is not monotonic towards higher harmonic orders, and at some point(s) R$_{i1}$ is higher than R$_{(i+1)1}$. The positions of these local minima (dips) vary from star to star, and do not seem to correlate with the pulsation periods. The sequences of amplitude ratios were investigated in RRL stars too \citep{Benko-2016}, but there, a slight correlation was detected between the periods and the positions and depths of the dips. 

%%%%%%%%%%%%%%%%%%%%%%%%%%%%%%%%%%%%%%%%%%%%%%%%%%%

\subsubsection{Anomalous Cepheids}
\label{subsec:ACs}

In the sample we found five ACs: EPIC 202862302 from Campaign 2, EPIC 218128117 from Campaign 7, and EPIC 246015642, 246385425 and 246333644 from Campaign 12. Their phased light curves are shown in Figure~\ref{fig:blh-ac} on the left panel with purple color. In the cases of EPIC 218128117 and 246385425 the decisive feature regarding classification is the bump on their descending branch, which is not present in any classical Cepheid (DCEP) with similar pulsation periods, and their positions in the Fourier parameter space (see Figure~\ref{fig:Fourier}). The stars EPIC 246333644 and 246385425 are investigated in further detail, because they show some scatter in their light curves, but it was determined that the scatter is a consequence of the photon noise set by the light-gathering power of the telescope, and not a physical phenomena in the stars. We do not detect a Blazhko-like modulation in these stars. 

In Figure~\ref{fig:blh-ac}, the right panel shows the Observed--Calculated (O--C) diagrams for the pulsation periods of the stars for the duration of the \textit{K2} data set. The O--C values were computed from the maxima of the pulsation cycles with a template fitting method. These values show a low scatter in the range of a few minutes (the maximum is $\sim10$ minutes in EPIC 202862302). The pulsation amplitudes also vary within $\sim$5 mmag. At this low level we cannot decide whether these are instrumental or intrinsic variations, especially in the fainter stars where observational noise is significant (in the case of the star EPIC 46385425 it can reach 40 mmag).

\citet{Kovtyukh_2018} noticed a presence of H$_{\alpha}$ emission in the spectra of the star EPIC 202862302, which was classified as a T2C previously. The measured metallicity would be compatible with an AC ([Fe/H]=$-1.94$ dex (see Tables~\ref{tab:literature} and \ref{tab:literature_continued}), as they are described in \citet{Fiorentino_2012}. Our new classification combined with detection of H$_{\alpha}$ emission by \citet{Kovtyukh_2018} would make this the first known AC showing such a feature. 

For the stars EPIC 218128117 and 246385425 the metallicity given in the literature (see Table~\ref{tab:literature} and \ref{tab:literature_continued}) is from the \textit{Gaia} DR2 \citep{2018AA...616A...1G}, where these stars were identified as DCEPs. The metallicity was derived from the Fourier parameters of the \textit{G}-band light curves. If these stars were DCEPs, then the transformation equation between the Fourier parameters and the [Fe/H] would be valid, but for ACs there is no known correlation (only few ACs have their metallicity measured from spectra). As a consequence, these metallicity data should be verified with direct spectroscopic measurements.

%%%%%%%%%%%%%%%%%%%%%%%%%%%%%%%%%%%%%%%%

\subsubsection{BL Herculis stars}
\label{subsec:BLHs}

The BLH stars from the \textit{K2} data sets are shown in Figure~\ref{fig:blh-ac} with green color. The shapes of the light curves are very typical for the pulsation periods of these types. No period doubling or any other period or amplitude change is detectable in the \textit{K2} light curves or in the O--C diagrams.

V839 Sgr or EPIC 215881928 in \textit{K2} was proposed as an RRL type variable, but the light curve shape and the period of its pulsation of 1.835 days confirm that it is a T2C, BLH subtype. In the \textit{K2} mission, 43 pulsation cycles of V389 Sgr were observed. In Figure~\ref{fig:blh-ac}, we see the phase folded light curve in the left panel, the whole measured light curve in the middle panel and the O--C diagram in the right panel. Neither one of the panels show any period or amplitude change in the star EPIC 215881928.

V527 Sgr or EPIC 217235287 in \textit{K2} was observed a few times in the past, by \citet{Kwee_1984}, by the SuperWASP survey\footnote{\url{https://wasp.cerit-sc.cz/form}}, and by members of the American Association of Variable Star Observers (AAVSO\footnote{\url{https://www.aavso.org/}}), for example. The International Variable Star Index (VSX\footnote{\url{https://www.aavso.org/vsx/}}) database has a remark stating that this star changed its period since the first time it was observed by \citet{Uitterdijk_1935}, who calculated it to be P=1.258956 days. The period measured from the \textit{K2} light curve is 1.2588231 days (see Table~\ref{tab:fou}). While this difference in the measured periods is detectable, and it can be traced to the first observations of this star, in the time span of the \textit{K2} observational cycle EPIC 217235287 did not show any period change. The O--C diagram is flat, see Figure~\ref{fig:blh-ac}, right panel.

%%%%%%%%%%%%%%%%%%%%%%%%%%%%%%%%%%%%%%%%%%%%%%%%%

\subsubsection{W Virginis stars}
\label{subsec:WVirs}

The photometric precision and the high cadence are the key to reveal tiny instabilities in the pulsation. Period doubling and the seemingly irregular cycle-to-cycle changes are known phenomena in T2Cs. These phenomena are common in longer period variables, but rare at the short period end of the T2Cs \citep{Catelan_book_2015}. 

The observed light curves of the five WVir stars in the \textit{K2} observing cycles are shown in Figure~\ref{fig:wvir} on the left panel, while the middle panel shows the phase folded light curves. The green color of the light curve changes with time, so when it is phase folded, the changes in the amplitude can be traced back to observational time. These light curves are studied in detail using the Fourier decomposition method. The Fourier spectra of these stars are shown in Figure~\ref{fig:wvir} on the right panel. The residual Fourier spectra have the fundamental frequency ($f_1$) and its harmonics ($f_2=2 f_{1}$, $f_3=3 f_{1}$,$f_4=4 f_{1}$ and $f_5=5 f_{1}$) up to the the fourth order removed. The positions of the $f_1$, $f_2$, $f_3$, $f_4$ and $f_5$ components are marked with vertical green dashed lines. 

The O--C analysis was not applied to the WVir stars, since the total observed pulsation periods cover only a few cycles, so it would not give a meaningful result. In Figure~\ref{fig:wvir} one can see that the WVir stars show cycle-to-cycle variation. The difference between the amplitudes of the consecutive cycles is in the range of a few mmag. 

Only one star, EPIC 210622262 shows an alternation of the even and odd cycles, the typical sign of period doubling (upper panels in Figure~\ref{fig:wvir}). The alternation is in the range of $\sim8$ mmag. Period doubling is also seen in the residual spectra in the form of a sub-harmonic peaks near $0.5f_1$ and $1.5f_1$. We note that only four cycles are measured in the K2 mission for this star, but since period doubling is expected in WVir stars above P~>~16 days, it is very likely that we see this phenomenon here, as well. Due to the short time span of the measurement, the period-4 behaviour can not be confirmed (4$\times$16.6=66.6 day, which is too close to the length of the observing run), but it is highly likely to be present in this star. After phasing the data with 2$\times$16.6524 day period, the minima and maxima seem to split into two values. 

The star EPIC 247445057 (VZ Tau) was found to be a member of the open cluster Platais 4 by \citet{Zejda_2012}. Since VZ Tau is pulsating in a regular manner as a WVir type star, it would be worth investigating if it is truly an open cluster member, since a WVir star should belong to an old population of stellar objects. In Figure~\ref{fig:wvir}, on the left panel, the light curve of this star shows changes in the shape of the maximum varying from one pulsation to the other. Again, this cannot be further investigated due to the shortness time span of the observations in each \textit{K2} observational cycle, but would be worth investigating further in detail.

%%%%%%%%%%%%%%%%%%%%%%%%%%%%%%%%%%%%%%%%%%%%%%%%

\section{Physical Parameters}
\label{sec:physical_param}

%%%%%%%%%%%%%%%%%%%%%%%%%%%%%%%%%%%%%%%%%%%%%%%%

In this section, we explore different techniques that can tell us more about the physical parameters of the stars in our \textit{K2} sample. We calculate the effective temperatures and luminosities (see Subsection \ref{subsec:SED}). The evolutionary status of the stars is examined in Subsection~\ref{subsec:hrd}. We use data from the \textit{Gaia} space observatory to plot the color-magnitude diagram (more in Subsection~\ref{subsec:CMD}). In Subsection~\ref{subsec:PR_relation}, we calculate the radii of the sample stars. We examine if any of these methods would be more suitable for the classification than the Fourier parameters derived from the light curves in combination with the light curve shapes. 

%%%%%%%%%%%%%%%%%%%%%%%%%%%%%%%%%%%%%%%%%%%%%%%%%

\subsection{Spectral Energy Distribution}
\label{subsec:SED}

The physical parameters of individual stars were determined using the program
\texttt{More of DUSTY}, MoD \citep{2012A&A...543A..36G}, an extension of the dust radiative transfer code DUSTY \citep{1999ascl.soft11001I}. For a given distance, reddening, and stellar model atmosphere the program determines the best-fitting luminosity by comparing the calculated spectral energy distribution (SED) to a set of observed magnitudes with errors.
By testing different model atmospheres the best-fitting photometric effective temperature and luminosity is determined. For lower metallicities the calculated luminosities get larger. The effect is small: $+3\%$ for 1 dex in [Fe/H].
MARCS model atmospheres are used \citep{Gustafsson_MARCS}, 
for $\log g$ = 2.0, one solar mass, and solar metallicity.
The effect of using non-solar metallicty models on the best-fitting luminosity is negligible compared to the errors
in distance and reddening.
None of the stars show evidence for any significant infrared excess and the dust optical depth is set to zero.

Distances are taken from \citet{BJ21} which are based on {\it Gaia} Early Data Release 3 (EDR3) data \citep{2021A&A...649A...1G}. For a detailed discussion on the precision of the distance determination see Section~\ref{sec:PL_relation}.
The interstellar reddening is determined from the 3D reddening map of \citet{Green2019}\footnote{\url{http://argonaut.skymaps.info} the `Bayestar19' dataset.}
that is based on {\it Gaia} DR2, 2MASS, and Pan-STARRS~1 data. For EPIC246015642 this value is not available and the 
3D reddening map of \cite{Lallement18}\footnote{\url{https://stilism.obspm.fr/}} is used.
%For a given galactic longitude, latitude and distance, the tool returns the value of $E(B-V)$ and error, and the distance to which these values refer.
%If this distance is smaller than the input distance the returned value for the reddening is a lower limit.
%In these cases a simple estimate of the reddening at the distance of the cepheid is made.
%A second reddening value is queried at a distance 0.75 times the maximum distance available in the grid in that direction.
%Based on this the first derivative (with error bar) is determined and the reddening at the distance of the target estimated.
%
%The error bar returned by STILISM is added in quadrature with the error due to a 1 degree change in $l$ and $b$, and a 5\% error in distance.

The data for the SED fit are collected from a plethora of sources.
In the UV we used data from the Galaxy Evolution Explorer (GALEX, \citealt{2017ApJS..230...24B}).
The visible and infrared parts of the spectra were covered with observations from
\textit{Gaia} EDR3 (the $G, G_\text{BP}$, and $G_\text{RP}$  magnitudes;  \citealt{2021A&A...649A...1G}),
the Tycho catalog \citep{2000A&A...355L..27H},
the 2MASS catalog \citep{2006AJ....131.1163S, 2003yCat.2246....0C},
the DENIS catalog \citep{Epchtein99}\footnote{Vizier catalog B/denis/denis},
the UKIDSS survey results \citep{2008MNRAS.391..136L},
the Vista Hemisphere Survey (VHS; \citealt{McMahon13})\footnote{Vizier catalog II/359}, and
the AllWISE catalog \citep{Wright10}\footnote{Vizier catalog II/328}.

\begin{figure*}
\begin{minipage}{0.40\textwidth}
\resizebox{\hsize}{!}{\includegraphics[angle=0]{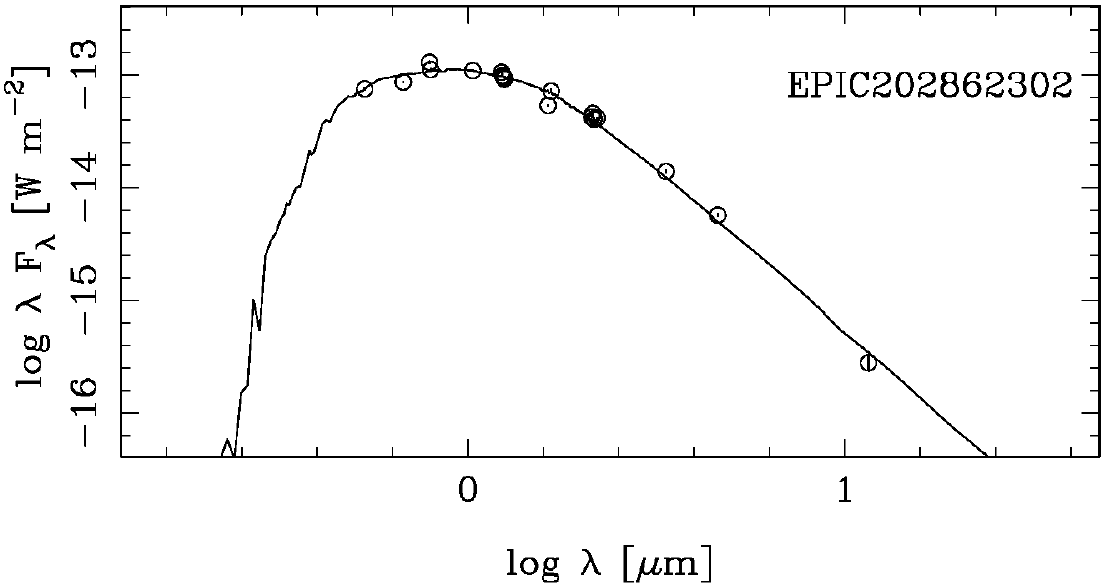}} 
\end{minipage}
\begin{minipage}{0.40\textwidth}
\resizebox{\hsize}{!}{\includegraphics[angle=0]{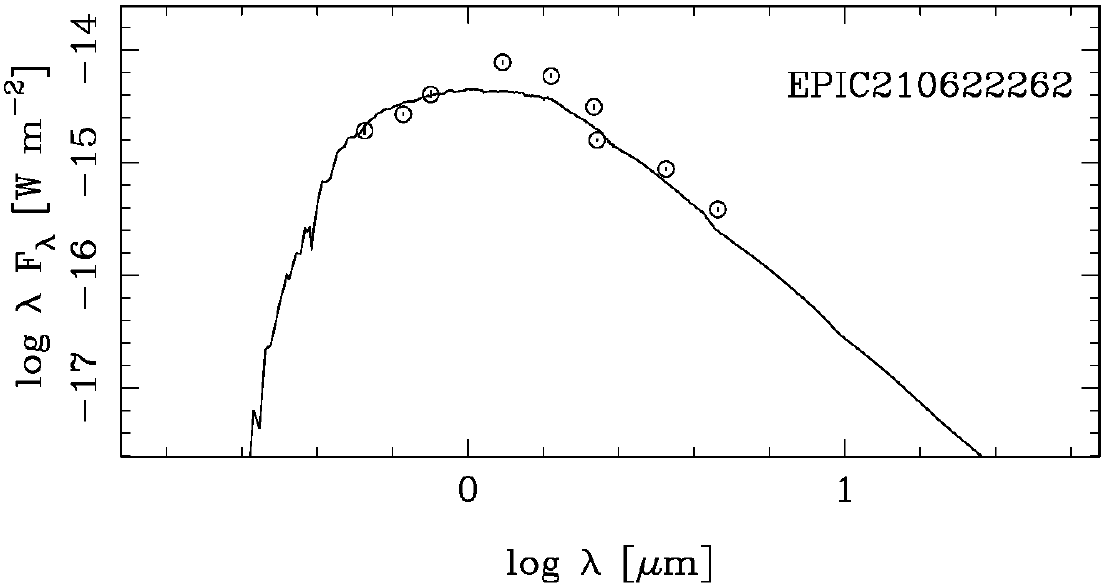}} 
\end{minipage}
 
\begin{minipage}{0.40\textwidth}
\resizebox{\hsize}{!}{\includegraphics[angle=0]{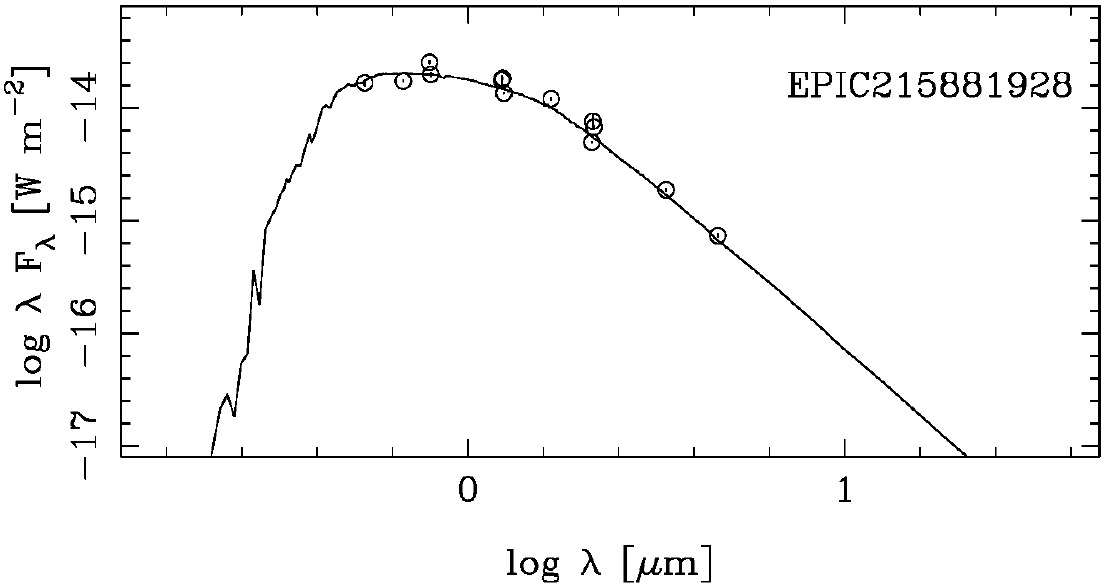}} 
\end{minipage}
\begin{minipage}{0.40\textwidth}
\resizebox{\hsize}{!}{\includegraphics[angle=0]{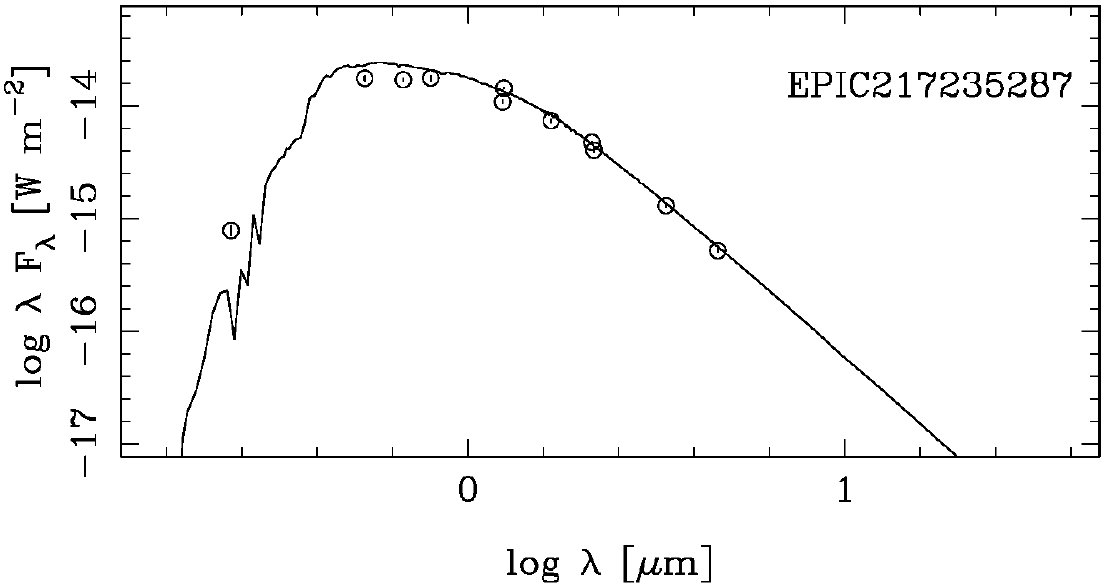}} 
\end{minipage}
 
\begin{minipage}{0.40\textwidth}
\resizebox{\hsize}{!}{\includegraphics[angle=0]{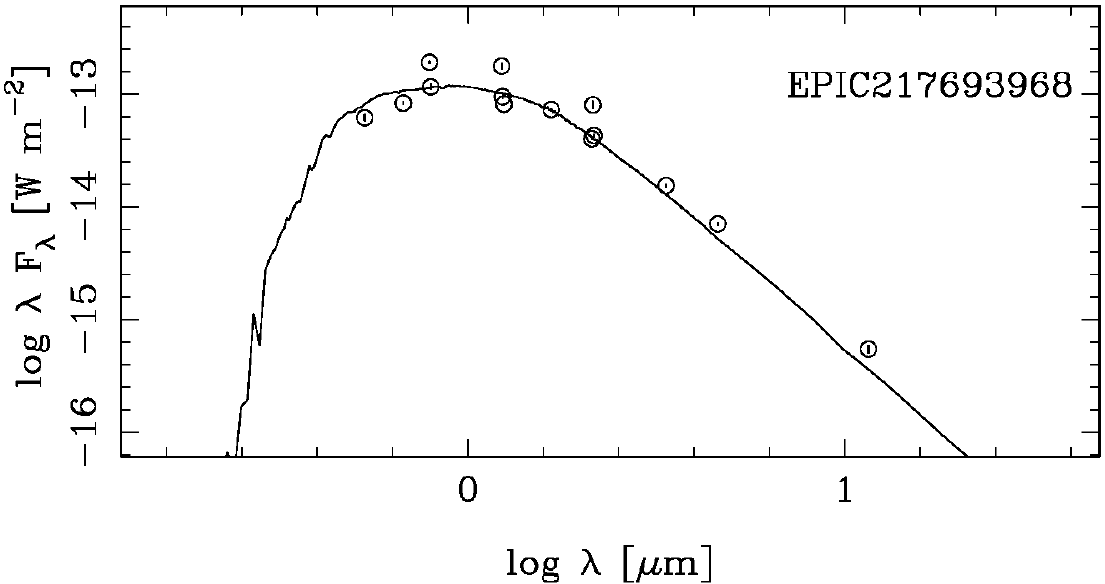}} 
\end{minipage}
\begin{minipage}{0.40\textwidth}
\resizebox{\hsize}{!}{\includegraphics[angle=0]{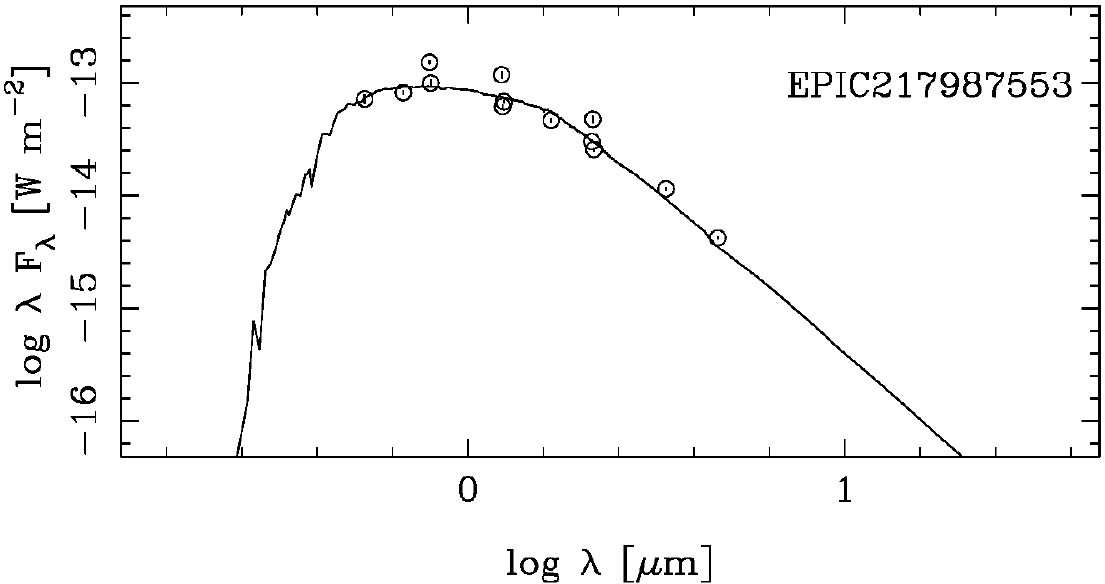}} 
\end{minipage}
 
\begin{minipage}{0.40\textwidth}
\resizebox{\hsize}{!}{\includegraphics[angle=0]{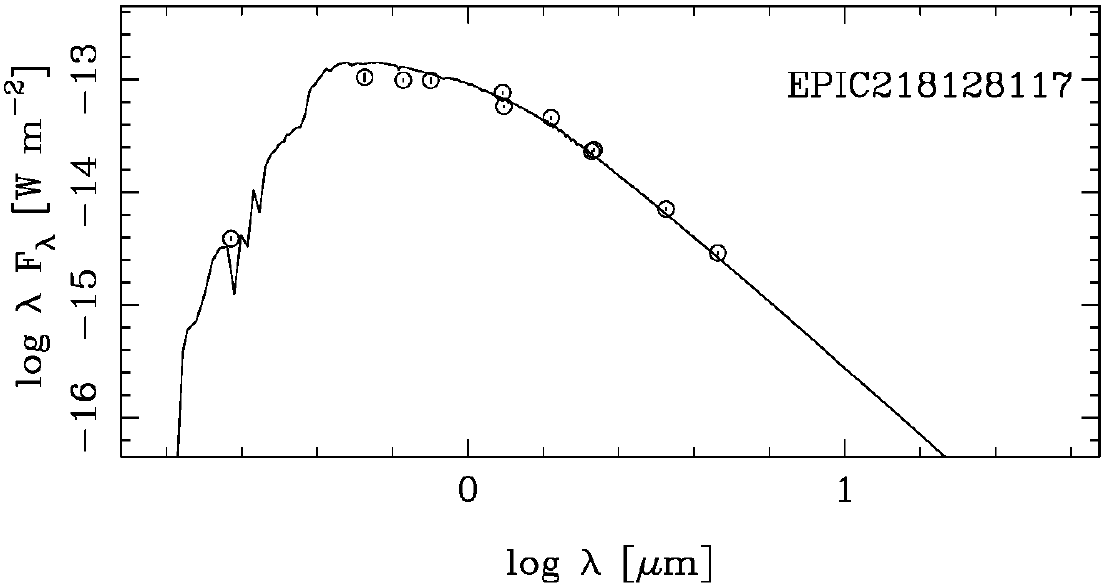}} 
\end{minipage}
\begin{minipage}{0.40\textwidth}
\resizebox{\hsize}{!}{\includegraphics[angle=0]{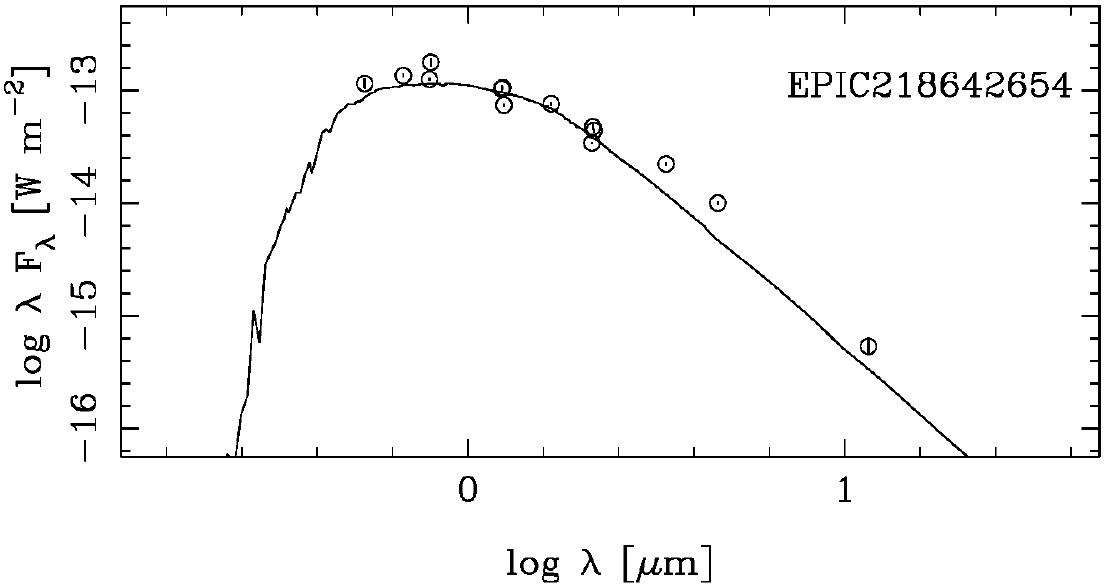}} 
\end{minipage}
 
\begin{minipage}{0.40\textwidth}
\resizebox{\hsize}{!}{\includegraphics[angle=0]{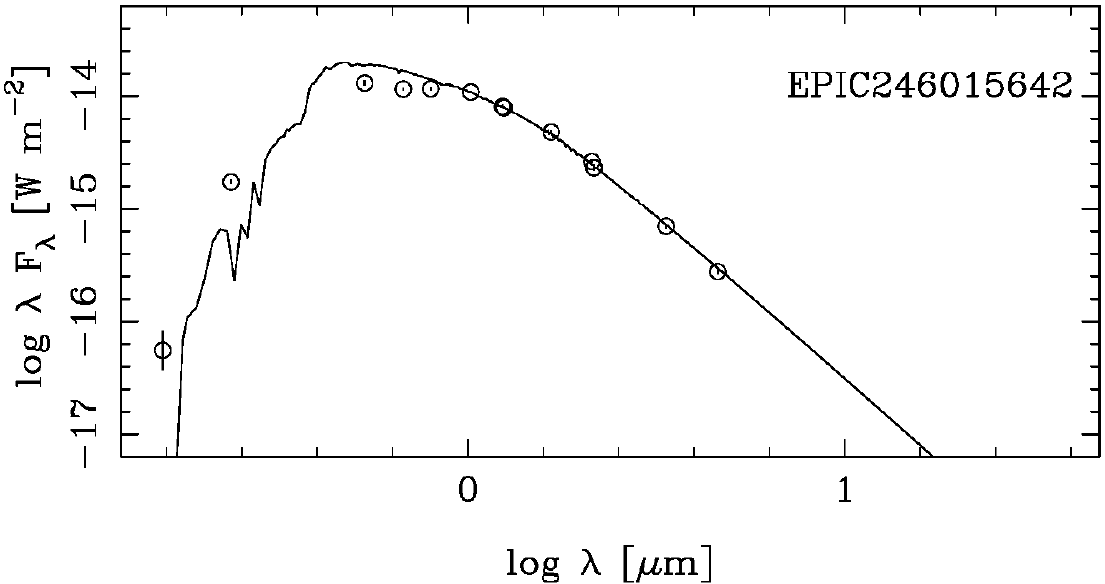}} 
\end{minipage}
\begin{minipage}{0.40\textwidth}
\resizebox{\hsize}{!}{\includegraphics[angle=0]{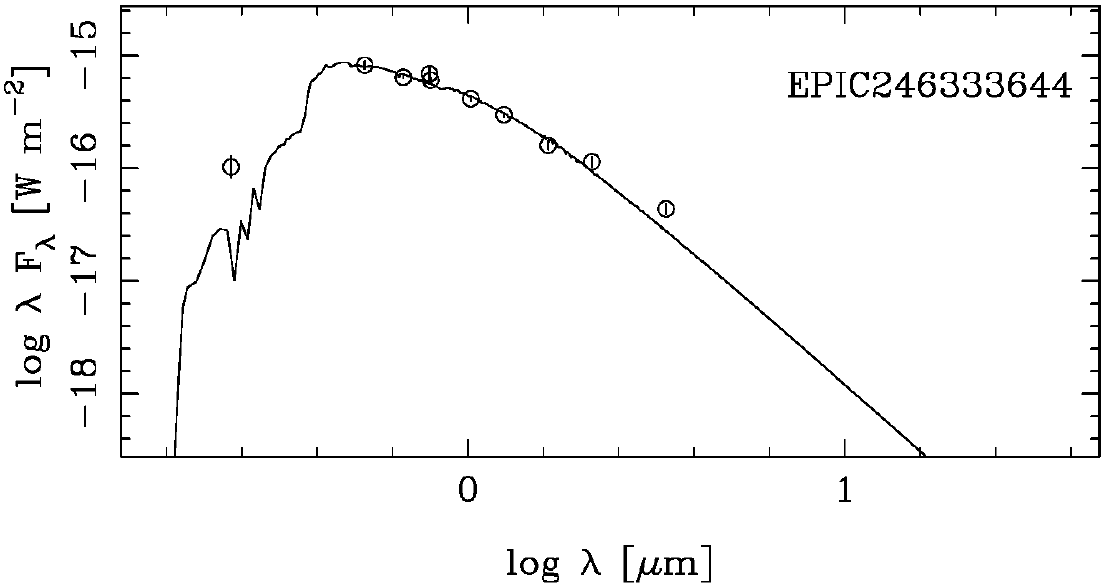}} 
\end{minipage}

\begin{minipage}{0.40\textwidth}
\resizebox{\hsize}{!}{\includegraphics[angle=0]{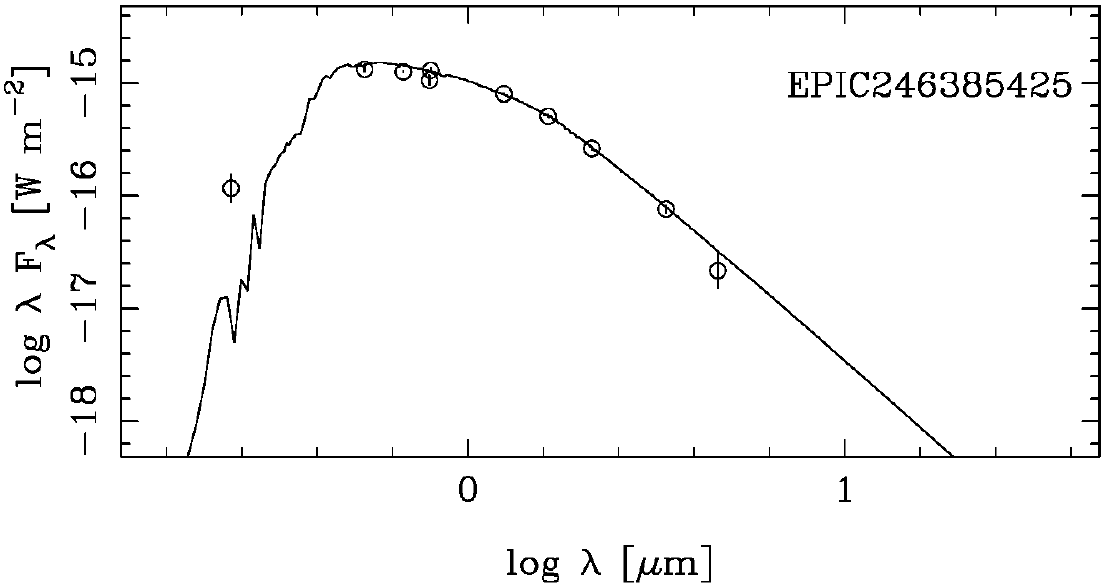}} 
\end{minipage}
\begin{minipage}{0.40\textwidth}
\resizebox{\hsize}{!}{\includegraphics[angle=0]{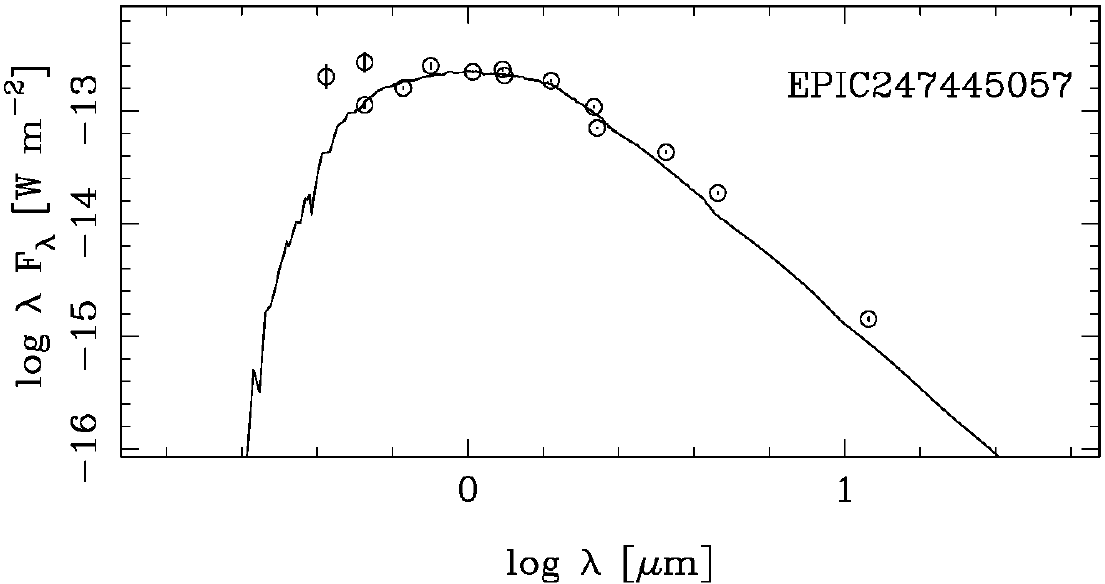}} 
\end{minipage}

\caption{Model fits (the solid lines) to the observed photometry (points with error bars). }
    \label{fig:SEDs}

\end{figure*}

The resulting best fits are shown in Figure~\ref{fig:SEDs}, and the parameters are summarised in Table~\ref{Tab:SED}. Columns~2, 3, 4, and 5 give the {\it Gaia} EDR3 source ID, parallax, Goodness of Fit (GoF), and Renormalised Unit Weight Error (RUWE). Three of the GoF values are large ($>$6), but all RUWEs are smaller than 1.4 which is considered to be cutoff for problematic astrometric solutions (where a problematic astrometric solution could be an indicator of a binary system). Column~6 lists the distance from \citet{BJ21}, and Column~7 is the interstellar reddening. The last two columns list the results from the fitting, namely the best fitting effective temperature and luminosity. The error in $T_{\rm eff}$ is about 125~K, while the error in the luminosity is the internal fitting error for the adopted distance.

\begin{table*}
\centering
\caption{Parameters from \it Gaia \rm and results from the fitting.}
\begin{tabular}{lrrllllll}
EPIC         &  source ID           &  parallax           & GoF   & RUWE   &  Distance        &  $A_{\rm V}$ & $T_{\rm eff}$ &  $L$    \\ 
             &      ERD3                & (mas)               &       &        &  (pc)            & (mag)      &  (K)       & (L$_{\sun}$)  \\ 
\hline
T2Cs  & & & & & & & & \\
\hline
210622262 & 47299585774090112 & -0.1003 $\pm$ 0.0780 & 0.6217 & 1.027 & 7383 $^{+2101 }_{-1728 }$ & 1.24 & 5000 &   19.33   $\pm$    1.99  \\
 & & & & & & & & \\[-1em]
217235287 & 4082823506754098432 & 0.0639 $\pm$ 0.0329 & 1.2925 & 1.064 & 8352 $^{+2414 }_{-1505 }$ & 0.43 & 6250 &   92.83   $\pm$    3.70  \\
 & & & & & & & & \\[-1em]
215881928 & 4075390224042082688 & 0.1274 $\pm$ 0.0269 & 2.0615 & 1.102 & 6088 $^{+1321 }_{-791 }$ & 0.50 & 5625 &   44.50   $\pm$    1.66  \\
 & & & & & & & & \\[-1em]
217987553 & 4085507620804499328 & 0.1495 $\pm$ 0.0199 & 0.8269 & 1.039 & 5297 $^{+570 }_{-549 }$ & 0.47 & 5250 &  150.68   $\pm$    6.48  \\
 & & & & & & & & \\[-1em]
218642654 & 4087335043492541696 & 0.0505 $\pm$ 0.0186 & 7.9334 & 1.387 & 11574 $^{+1806 }_{-1735 }$ & 0.71 & 5500 &  1085.55   $\pm$   86.12  \\
 & & & & & & & & \\[-1em]
217693968 & 4085983537561699584 & 0.1929 $\pm$ 0.0194 & 6.1186 & 1.289 & 4327 $^{+452 }_{-359 }$ & 1.15 & 5750 &  191.96   $\pm$   11.51  \\
 & & & & & & & & \\[-1em]
247445057 & 3415206707852656384 & 0.2708 $\pm$ 0.0225 & 8.1617 & 1.376 & 3499 $^{+236 }_{-252 }$ & 1.15 & 5125 &  214.78   $\pm$   11.49  \\
 & & & & & & & & \\[-1em]
\hline
ACs  & & & & & & & & \\
\hline
202862302 & 6044434301761335808 & 0.1830 $\pm$ 0.0180 & 0.32 & 1.013 & 4518 $^{+323 }_{-431 }$ & 1.18 & 5750 &  200.66   $\pm$    5.86  \\
 & & & & & & & & \\[-1em]
218128117 & 6869460685678439040 & 0.0967 $\pm$ 0.0201 & 3.9358 & 1.198 & 7457 $^{+1115 }_{-957 }$ & 0.31 & 6500 &  390.41   $\pm$   14.71  \\
 & & & & & & & & \\[-1em]
246015642 & 2435741447518507392 & 0.0453 $\pm$ 0.0331 & 3.4579 & 1.162 & 8353 $^{+2643 }_{-1422 }$ & 0.10 & 6500 &   54.36   $\pm$    2.98  \\
 & & & & & & & & \\[-1em]
246385425 & 2638680812622984960 & 0.1101 $\pm$ 0.0995 & -0.3849 & 0.980 & 4504 $^{+1789 }_{-1098 }$ & 0.16 & 6000 &    1.36   $\pm$    0.03  \\
 & & & & & & & & \\[-1em]
246333644 & 2637645489281432832 & -0.0774 $\pm$ 0.1914 & 0.9908 & 1.049 & 3722 $^{+1664 }_{-1168 }$ & 0.16 & 6750 &    0.47   $\pm$    0.02  \\
 & & & & & & & & \\[-1em]
\hline

\end{tabular} 

\label{Tab:SED}
\end{table*}

%%%%%%%%%%%%%%%%%%%%%%%%%%%%%%%%%%%%%%%%%%%%%%%

\subsection{The Hertzsprung--Russell Diagram}
\label{subsec:hrd}

\begin{figure}
\centering
    \includegraphics[width=.45\textwidth]{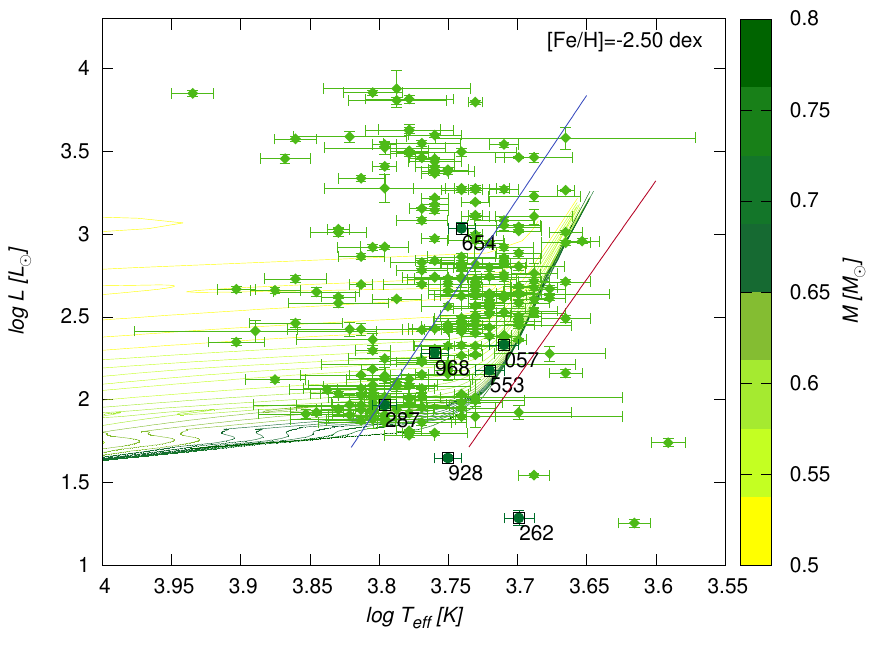}\hfill
    \includegraphics[width=.45\textwidth]{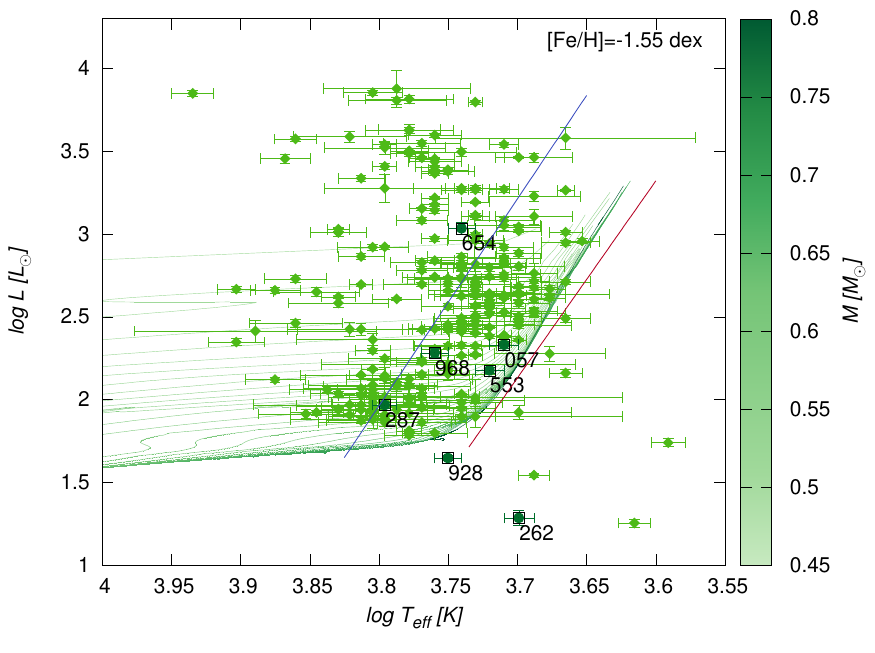}\hfill
    \includegraphics[width=.45\textwidth]{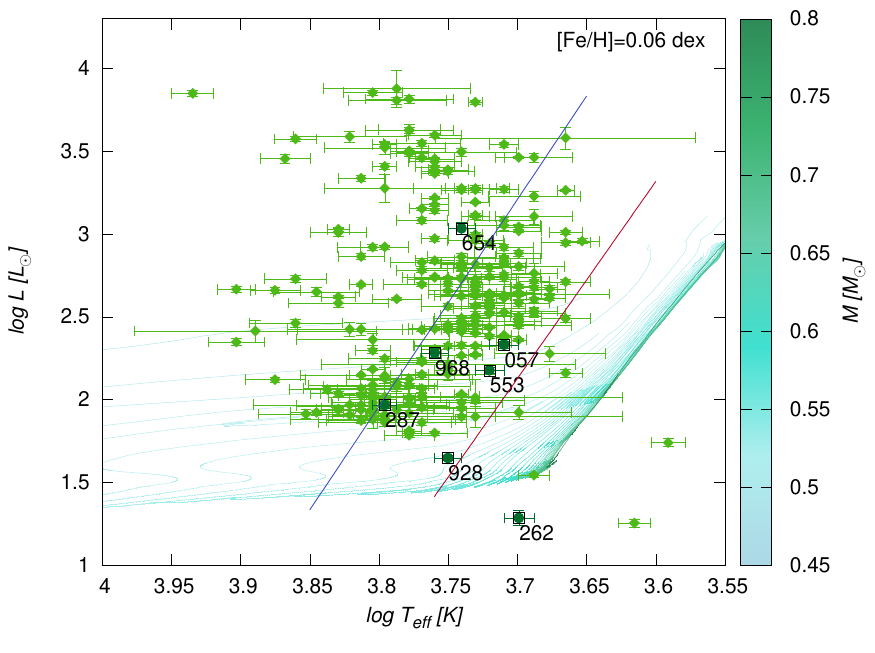}
\caption{HRD: BaSTI models for low-mass stars evolving from the Horizontal Branch (HB) representing the T2C evolutionary tracks. Upper panel: [Fe/H]=$-2.50$, middle panel: [Fe/H]=$-1.55$, bottom panel: [Fe/H]= $0.06$. The green crosses are T2Cs from the SMC and LMC taken from the \citet{Groenewegen_2017a}. The colored bars show the range of masses in solar masses from 0.45 to 0.8 $M_{\sun}$.}
\label{fig:t2c_hrd}
\end{figure}

\begin{figure*}
\centering
    \begin{minipage}[b]{.45\textwidth}
        \includegraphics{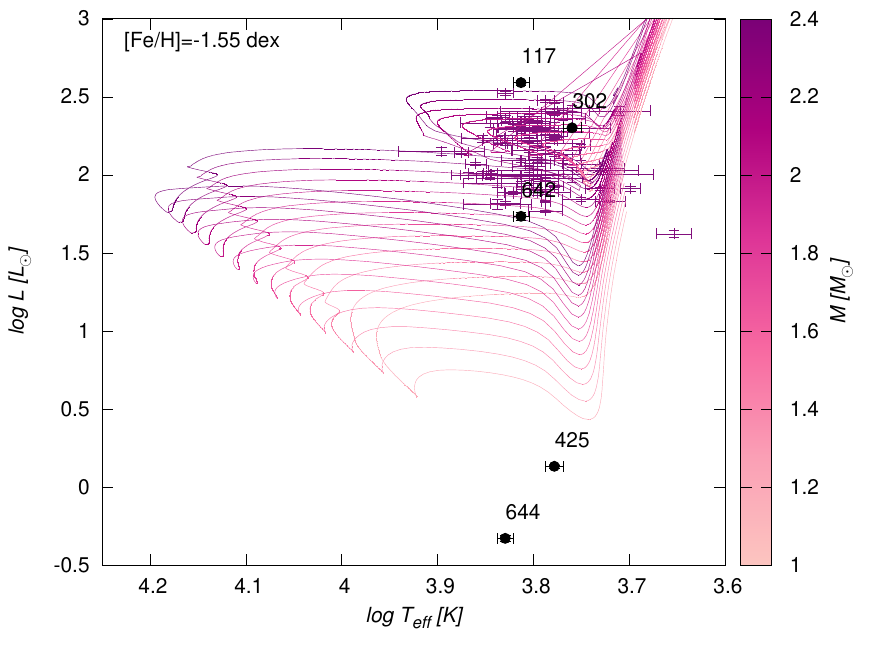}
        \end{minipage}
    \qquad
    \begin{minipage}[b]{.45\textwidth}
        \includegraphics{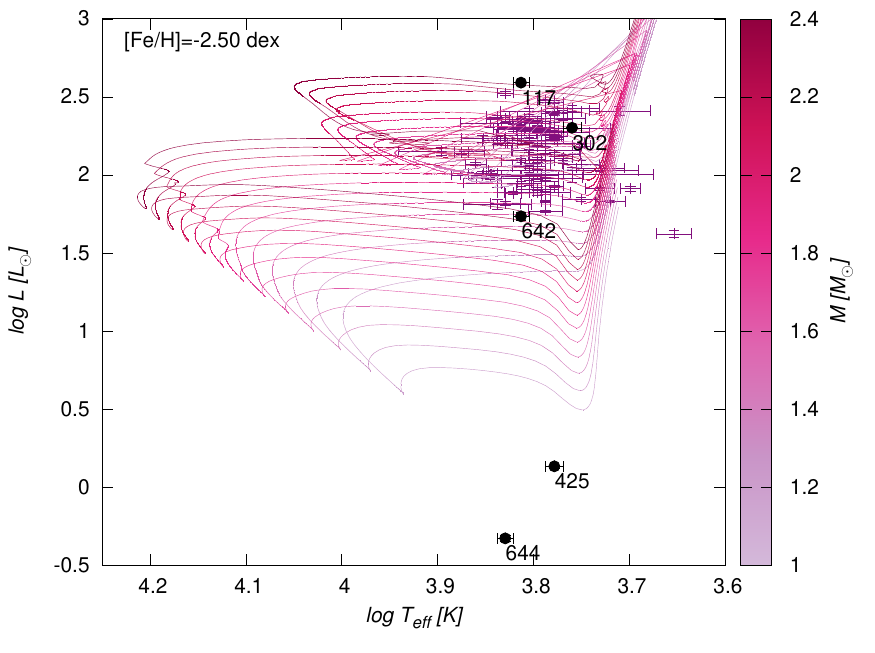}
    \end{minipage}
 \caption{HRD: BaSTI models for single evolutionary models of ACs - on the left side [Fe/H]=$-1.55$, and on the right side [Fe/H]= $-2.50$. Stars 644 and 425 are ACs which are among the least bright ones, so their luminosity value is probably unreliable due to their parallax measurement error. The lilac crosses are ACs from the SMC and LMC taken from the \citet{Groenewegen_2017a}. The mass range, from 2.4 $M_{\sun}$ down to 1.0 $M_{\sun}$ with steps of 0.1 $M_{\sun}$, is indicated by the change in the color bar shown on the right hand side of the plot.}
\label{fig:ac_hrd}
\end{figure*}

The Hertzsprung--Russell Diagram (HRD) is constructed using the $L$ and $T_{\textrm{eff}}$ values for the stars we have examined (see Table~\ref{tab:stars_in_K2}). Figures~\ref{fig:t2c_hrd} and \ref{fig:ac_hrd} show the place of our program stars together with the T2Cs and ACs taken from \citet{Groenewegen_2017a} for the Small Magellanic Cloud (SMC) and LMC. The evolutionary models are from the BaSTI (Bag of Stellar Tracks and Isochrones\footnote{\url{http://basti-iac.oa-abruzzo.inaf.it/index.html}}) models \citep{2018ApJ...856..125H}. The use models with solar-scaled heavy metal elements, with no diffusion, and no mass-loss.

The literature data for $T_{\textrm{eff}}$, the surface gravity (log\,$g$), and metallicity ([Fe/H]) are given in Table~\ref{tab:literature}. From the seven T2Cs in our \textit{K2} sample, four have metallicity data available in the literature. Two of them, EPIC 217987553 and 217696968, have metallicities, [Fe/H]=$-0.804$ and $-0.567$, respectively, putting them in the slightly metal-poor category, while two more, EPIC 217235287 and 215881928, have metallicities of [Fe/H]=$-0.042$ and $-0.058$, making them stars with solar-like metallicities. These measurements motivated us to plot the BaSTI models with three different metallicities for T2Cs. In the case of the T2Cs the metallicities are [Fe/H]=$-2.50$, $-1.55$, and $+0.06$ dex. The horizontal branch (HB) models are shown with their masses ranging from 0.47 $M_{\sun}$ to 0.80 $M_{\sun}$. The ACs in Table~\ref{tab:literature} have metallicities that range from [Fe/H]=$-2.00$ to $-0.01$ which is higher than what the models from \citet{2006A&A...460..155F} predict (Z=$0.0001$ or [Fe/H]=$-2.36$), but are in good agreement with the models from \citet{Caputo_2004} in which the metallicities of ACs are Z=0.0004. The most reliable metallicities come from direct spectroscopic measurements, and in Tables~\ref{tab:literature} and \ref{tab:literature_continued} these are [Fe/H]=$-1.94$ for EPIC 202862302 from \citet{2019A&A...628A..94A}; and [Fe/H]=$-1.513$ and [Fe/H]=$-2.00$ for EPIC 218128117 and for EPIC 2460155642, respectively, from \citet{2020ApJS..247...28H}. The other metallicity and stellar parameter data are derived from photometry, astrometry, and other available data form catalogs, and the estimates made from them. The large differences in the metallicity data for each star could be the consequence of different methods used to determine the metallicities.

The models with the metallicities of $-2.50$ and $-1.55$ cover most of the T2Cs from the LMC and SMC, as well as the stars from our sample, EPIC 217235287, 217987553, 218642654, 247445057, and 217693968 shown in Figure~\ref{fig:t2c_hrd}. This is in agreement with the findings of \citet{Groenewegen_2017a} and \citet{2020A&A...644A..96B}. In the case of EPIC 215881928 only the solar-like metallicity could reproduce its position on the HRD, which is in agreement with the metallicity from the literature ([Fe/H]=$-0.058$). EPIC 210622262 is outside the limits of the evolutionary models, but the distance estimation, and thus the $L$ and $T_{\textrm{eff}}$ are uncertain, which could result in its current position on the HRD being inaccurate. EPIC 217693968 has a mass estimation from \citet{Huber_2016} of M$^{c}$=0.963 M$_{\odot}$, but on all the evolutionary model figures (see Figure~\ref{fig:t2c_hrd}), only lower mass models, between 0.5 and 0.6 $M_{\sun}$, cross the position of this star, making the previous mass estimation too high.

The star EPIC 210622262, which was identified as a WVir type star that is showing a PD effect in its light curve, is outside the $-2.50$ and $-1.55$ metallicity evolutionary tracks, but is close to the solar-like ($+0.06$) metallicity model of 0.8 $M_{\sun}$ (together with some other stars from the \citet{Groenewegen_2017a} sample). In  \citet{Smolec_2016}, the PD phenomenon is explained in detail for the T2Cs with masses from 0.6 $M_{\sun}$ up to 0.85 $M_{\sun}$, however the metallicity of all the models is low. In this way, the models from \citet{Smolec_2016} do not give us an explanation of the PD phenomena in the solar-like metallicity star of 0.8 $M_{\sun}$. 

\begin{figure}
\centering
    \includegraphics{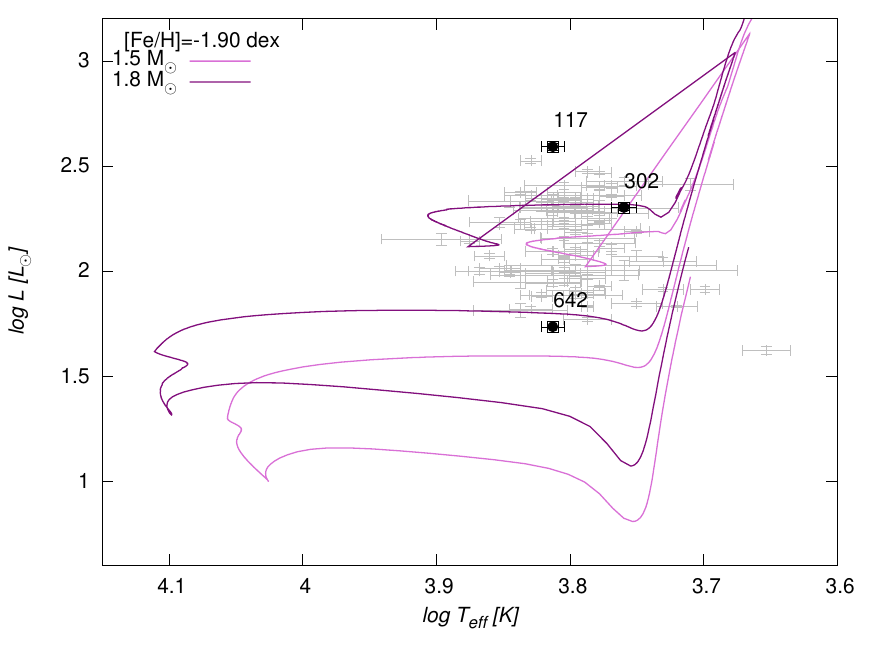}
    \caption{HRD: BaSTI models for single evolution of the AC EPIC 202862302. The purple line is a model of mass M=1.8 $M_{\odot}$ and the pink is for mass M=1.5$M_{\odot}$. The metallicity of the model is [Fe/H]=-1.90 dex, which is in good agreement of the measured metallicity of EPIC 202862302, [Fe/H]=-1.94 dex. The grey crosses are ACs from the SMC and LMC taken from the \citet{Groenewegen_2017a}.}
\label{fig:302_HRD}
\end{figure}

The ACs EPIC 216015642, 202862302, and 218128177 can be fitted well with the evolutionary models with metallicities of $-1.55$ and $-2.50$ as seen in Figure~\ref{fig:ac_hrd}. In the case of EPIC 246333644 and 246385425 we see that they are outside the modeled HRD region. They are among the faintest stars in the sample, and due to the distance uncertainty their $L$ was probably underestimated. 
The cases of EPIC 218128117 and 246385425 need further detailed investigation, since the available metallicity values in the literature (see Tables~\ref{tab:literature} and \ref{tab:literature_continued}) are $-0.01$ and $0.12$, respectively, which are too high for a typical AC. 
The problem with these data is that the \textit{Gaia} DR2 estimates are based on their previous classification as DCEP stars, so the metallicity estimated from the Fourier parameters uses the correlation between the Fourier parameters and the metallicity for the DCEP stars. \citet{Gautschy_2017} shows different scenarios for binary channels to produce ACs, but non of their models cover the luminosity range of EPIC 246333644 and 246385425 either. If ACs would come from a binary channel evolution, they would have gone through a rapid mass-loss phase \citep{Gautschy_2017}, but we do not see evidence of IR-excess in the SED around the observed ACs (see Figure~\ref{fig:SEDs}).

For the AC EPIC 202862302 \citet{Kovtyukh_2018} has measured a metallicity of [Fe/H]=$-1.94$ dex. Looking at the evolutionary models of metal-poor stars from the BaSTI database two masses could be fitted on the position where EPIC 202862302 is on the HRD for the metallicity of [Fe/H]=$-1.90$ dex: 1.5 and 1.8  $M_{\odot}$, as it can seen in Figure~\ref{fig:302_HRD}.

\begin{table*}
    \centering
	\caption{Effective temperatures ($T_{\textrm{eff}}$), surface gravity (log\,$g$), metallicities ([Fe/H]), masses (M$_{\odot}$) and radii (R$_{\odot}$) from the literature for the AC stars in the \textit{K2} sample. References:
	$^{a}$\citet{2015ApJ...811...30M};
	$^{b}$\citet{Huber_2016};
	$^{c}$\citet{2017AJ....154..259S};
	$^{d}$\citet{2018AA...616A...1G};
	$^{e}$\citet{Kovtyukh_2018};
	$^{f}$\citet{2018yCat.5153....0L};
	$^{g}$\citet{2018AJ....155...22S};
	$^{h}$\citet{2018ApJ...867..105T};
	$^{i}$\citet{2019A&A...628A..94A};
	$^{j}$\citet{2019AJ....158...93B};
	$^{k}$\citet{2019AJ....158..138S};
	$^{l}$\citet{2019ApJS..245...34X};
	$^{m}$\citet{2021A&A...649A...1G};
	$^{n}$\citet{2020ApJS..247...28H};
	$^{o}$\citet{2020MNRAS.495.3087L};
	$^{p}$\citet{2021A&A...651A..79B};
	$^{q}$\citet{2021MNRAS.506..150B};
	$^{r}$\citet{2022A&A...658A..91A}.}
	\label{tab:literature}
	\begin{tabular}{cccccc}
	ID & $T_{\textrm{eff}}^{\textrm{literature}}$ & log\,$g$ & [Fe/H] & $M$ & $R$ \\
	 & [K] & [cm/s$^2$] & [dex] & [M$_{\odot}$] & [R$_{\odot}$] \\
    \hline
    ACs  & & & & & \\
    \hline
    202862302 & 5428 $^{b}$ & 3.542 & -0.343 & 0.963 & 2.933 \\
    & & & & &\\[-1em] 
    & 4892.30 $\substack{+114.33 \\ -43.40}$ $^{d}$ & - & - & - & - \\
    & & & & &\\[-1em]
    & 5950$^{e}$ & 2.20 & -1.94 & - & - \\
    & & & & &\\[-1em]
    & 5432.32$^{i}$ & 2.212 & -1.17 & 2.528 & - \\
    & & & & &\\[-1em]
    & 4961.94$^{r}$ & 1.997 & -1.52 & 0.854 & - \\[5pt]
    & & & & &\\[-1em]
    218128117 & 7158$^{b}$ & 4.195 & -0.034 & 1.506 & 1.586 \\
    & & & & &\\[-1em]
    & 6345.00 $\substack{+463.25 \\ -325.69}$ $^{d}$ & - & -0.01 & - & - \\
    & & & & &\\[-1em]
    & 6345$^{h}$ & - & - & - & - \\
    & & & & &\\[-1em]
    & 6304.08$^{i}$& 2.665 & -1.513 & 1.854 & - \\
    & & & & &\\[-1em]
    & 6742 $\pm$ 291$^{j}$ & - & - & - & - \\
    & & & & &\\[-1em]
    & & & & & 8.959$^{k}$ \\
    & & & & &\\[-1em]
    & 7190$^{n}$ & 4.019 & -0.362 & 38.272 & 9.969 \\[5pt]
    & & & & & \\[-1em]
    246015642 & 6353$^{a}$ & - & -0.872, -0.742 & - & - \\
    & & & & &\\[-1em]
    & 6459.00 $\substack{+211.00 \\ -277.33}$ $^{d}$ & - & - & - & -\\
    & & & & &\\[-1em]
    & 6101.79$^{i}$ & 2.622 & -2.00 & 0.794 & - \\
    & & & & &\\[-1em]
    & - & - & - & 1.290$^{k, d}$ & 2.652 \\
    & & & & &\\[-1em]
    & 7098$^{n}$ & 4.039 & -0.162 & 2.902 & 2.689 \\
    & & & & &\\[-1em]
    & 5966.6$^{p}$ & 3.36 & -0.39, -0.76 & - & - \\[5pt]  
    & & & & &\\[-1em]
    246385425 & 6191$^{a}$ & - & -0.579, -0.548 & - & - \\
    & & & & &\\[-1em]
    & - & - & 0.12 $\substack{+0.4 \\ -0.5}$ $^{d}$ & - & - \\
    & & & & &\\[-1em]
    & 6346.22$^{i}$ & 4.366 & -0.919 & 0.900 & - \\
    & & & & &\\[-1em]
    & 6060.0 $\pm$ 285$^{k}$ & 4.846 & - & 1.130 & 0.664 \\
    & & & & &\\[-1em]
    & 5790.3$^{p}$ & 4.610 & -0.14, -0.15 & - & - \\
    & & & & &\\[-1em]
    & 6208.41$^{r}$ & 4.238 & -1.467 & 0.773 & - \\[5pt]
    & & & & &\\[-1em]
    246333644 & - & - & - & - & - \\[5pt]
    & & & & &\\[-1em]
    \hline
    \end{tabular}
\end{table*}
\begin{table*}
    \centering
	\caption{Same as Table \ref{tab:literature}, the T2C stars.}
	\label{tab:literature_continued}
	\begin{tabular}{cccccc}
	ID & $T_{\textrm{eff}}^{\textrm{literature}}$ & log\,$g$ & [Fe/H] & $M$ & $R$  \\
	 & [K] & [cm/s$^2$] & [dex] & [M$_{\odot}$] & [R$_{\odot}$] \\
    \hline
    T2Cs  & & & & &\\
    \hline
    210622262 & 5396$^{b}$ & 4.527 & -0.237 & 0.868 & 0.841 \\
    & & & & &\\[-1em]
    & 4349.70 $\substack{+193.55 \\ -94.70}$ $^{d}$ & - & - & - & - \\
    & & & & &\\[-1em]
    & 6955.06$^{f}$ & - & -0.095 & - & - \\
    & & & & &\\[-1em]
    & 5161.17 $\pm$ 6.72$^{l}$ & 1.512 $\pm$ 0.135 & -1.222 $\pm$ 0.089 & - & - \\
    & & & & &\\[-1em]
    & 5751$^{n}$ & 3.858 & -0.290 & 2.926 & 3.344 \\[5pt]
    & & & &\\[-1em]
    217235287 & 5816$^{b}$ & 4.159 & -0.042 & 1.025 & 1.366 \\
    & & & & &\\[-1em]
    & 5882.00 $\substack{+66.00 \\ -59.00}$ $^{d}$ & - & - & - & -\\
    & & & & &\\[-1em]
    & 5616.18$^{i}$ & 2.526 & -1.388 & 1.114 & - \\
    & & & & &\\[-1em]
    & - & - & - & - & 7.554$^{k}$ \\
    & & & & &\\[-1em]
    & 6150$^{n}$ & 4.062 & -0.360 & 27.644 & 8.077 \\[5pt]
    & & & & & \\[-1em]
    215881928 & 5966$^{b}$ & 4.222 & -0.058 & 1.062 & 1.300 \\
    & & & & &\\[-1em]
    & 5290.74 $\substack{+74.26 \\ -261.74}$ $^{d}$ & - & - & - & - \\
    & & & & &\\[-1em]
    & 5305.85$^{i}$ & 2.722 & -0.621 & 1.531 & - \\
    & & & & &\\[-1em]
    & 5724$^{n}$ & 3.621 & -0.899 & 5.519 & 5.989 \\
    & & & & &\\[-1em]
    & 5211.92$^{r}$ & 2.693 & -0.679 & 1.121 & - \\[5pt]
    & & & & &\\[-1em]
    217987553 & 5954$^{b}$ & 4.416 & -0.804 & - & - \\
    & 4985.00 $\substack{+390.00 \\ -383.70}$ $^{d}$ & - & - & - & - \\
    & & & & &\\[-1em]
    & 5475.06$^{i}$ & 2.451 & -0.798 & 2.255 & - \\
    & & & & &\\[-1em]   
    & - & - & - & - & 30.529$^{k}$ \\
    & & & & &\\[-1em]
    & 6029.06$^{r}$ & 4.013 & -0.560 & 0.914 & - \\[5pt]
    & & & & & \\[-1em]
    218642654 & 5265$^{b}$ & 2.902 & -0.773 & 0.907 & 4.900 \\
    & & & & & \\[-1em]
    & 4926.50 $\substack{+117.87 \\ -149.5}$ $^{d}$ & - & - & - & - \\
    & & & & & \\[-1em]
    & 4605 $\pm$ 114$^{j}$ & - & - & - & - \\
    & & & & & \\[-1em]
    & 5252.0 $\pm$ 402$^{k}$ & - & - & - & 45.220 \\
    & & & & & \\[-1em]
    & 4696.0 $\pm$ 99$^{o}$ & 2.63 $\pm$ 0.16 & - & - & - \\[5pt]
    & & & & & \\[-1em]
    217693968  & 4951$^{b}$ & 2.606 & -0.567 & 0.963 & 8.605 \\
    & & & & & \\[-1em]
    & 4537.61 $\substack{+313.39 \\ -106.61}$ $^{d}$ & - & - & - & - \\
    & & & & & \\[-1em]
    & - & - & - & - & 28.134$^{k}$ \\
    & & & & & \\[-1em]   
    & 5492$^{n}$ & 2.749 & -0.988 & 11.186 & 23.128 \\[5pt]
    & & & & & \\[-1em]
    247445057 & 6867.91$^{c}$ & - & - & - & - \\
    & & & & & \\[-1em]
    & 4211.00 $\substack{+77.74 \\ -104.29}$ $^{d}$ & - & - & - & - \\
    & & & & & \\[-1em]
    & 5419 $\pm$ 524$^{j}$ & - & - & - & - \\
    & & & & & \\[-1em]
    & 4837.0 $\pm$ 282$^{k}$ & - & - & - & 19.578 \\
    & & & & & \\[-1em]
    & 5000.0$^{m}$ & - & - & - & - \\
    & & & & & \\[-1em]
    & 5325$^{n}$ & 2.457 & -0.893 & 2.610 & 15.761 \\
    & & & & & \\[-1em]
    & 5164.49$^{q}$ & 2.236 & -0.327 & - & - \\[5pt]
    & & & & & \\[-1em]
    \hline
    \end{tabular}
\end{table*}

%%%%%%%%%%%%%%%%%%%%%%%%%%%%%%%%%%%%%%%%%%%%%%%%%%%%%%%%%%%%%%%%%%%%%%%%%%%%%%

\subsection{Color-magnitude diagram from \textit{Gaia} data}
\label{subsec:CMD}

\begin{figure*}
    \centering
    \includegraphics[width=\textwidth]{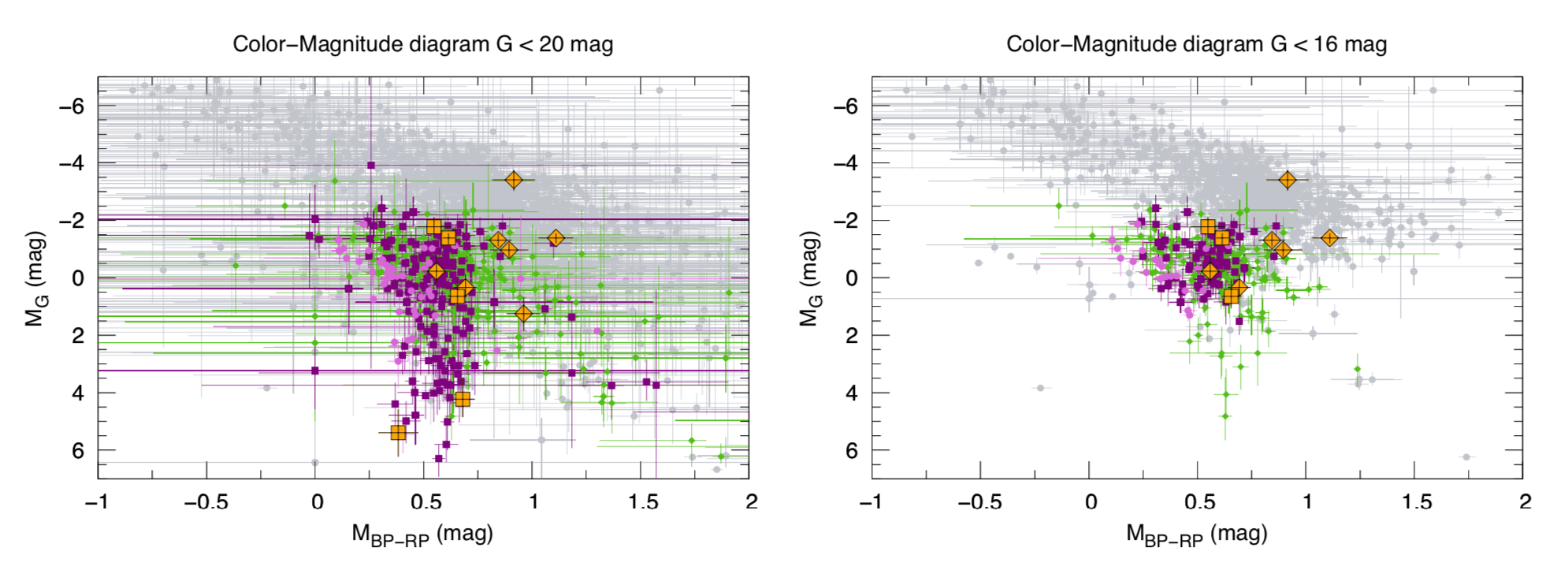}
    \caption{Color-magnitude diagrams for Cepheids with brightness limits of apparent magnitudes of 20 mag (left) and 16 mag (right) in Gaia $G$ band. Absolute brightness values were computed based on distances by \citet{Bailer-Jones_2018} considering interstellar dust. T2Cs (green) fundamental mode ACs (purple), overtone mode ACs (pink) and DCEPs (grey) were adopted from the catalogue by 
    \citet{2019A&A...625A..14R} and \citet{Soszynski_2017,sos2020}. \textit{K2} targets are marked with large orange symbols.}
    \label{fig:CMD}
\end{figure*}

The color data measured by \textit{Gaia} DR2 (in the blue filter (BP), red filter (RP) and the \textit{Gaia} filter (G)) were collected from the \textit{Gaia} database for the T2Cs and ACs in our \textit{K2} and for the known T2Cs, ACs and DCEPs. The list of known T2Cs (green), fundamental mode ACs (purple), overtone mode ACs (pink) and DCEPs (grey) were adopted from the catalogue by \citet{2019A&A...625A..14R} and \citet{Soszynski_2017,sos2020}. The absolute magnitudes were calculated using the distances from the \citet{Bailer-Jones_2018}. The interstellar dust was taken into account as it was detected by \textit{Gaia}.
Figure~\ref{fig:CMD} shows two sets of CMDs. On the left side the apparent G magnitude goes down to the 20 mag, and on the right side till 16 mag. The error bars on the colors are large, and the main source of the error is the reddening, and dust map inaccuracies. For stars with apparent magnitudes above 13-14 mag \citet{2021arXiv210907329M} the CMD is reliable in distinguishing between fundamental mode and overtone pulsation within one type of pulsating stars, but it is not too good for determining variable star classes (RRL vs ACEP vs DCEP vs T2C).

At this point the ACs and the T2Cs do not separate clearly, so the CMD should be used with caution when separating these subtypes of variable stars.

%%%%%%%%%%%%%%%%%%%%%%%%%%%%%%%%%%%%%%%%%%%%%%%%%%%%%%%%%%%%%%%%%%%%%5

\subsection{The Period-Radius Relation}
\label{subsec:PR_relation}

\begin{figure}
    \centering
    \includegraphics[width=\columnwidth]{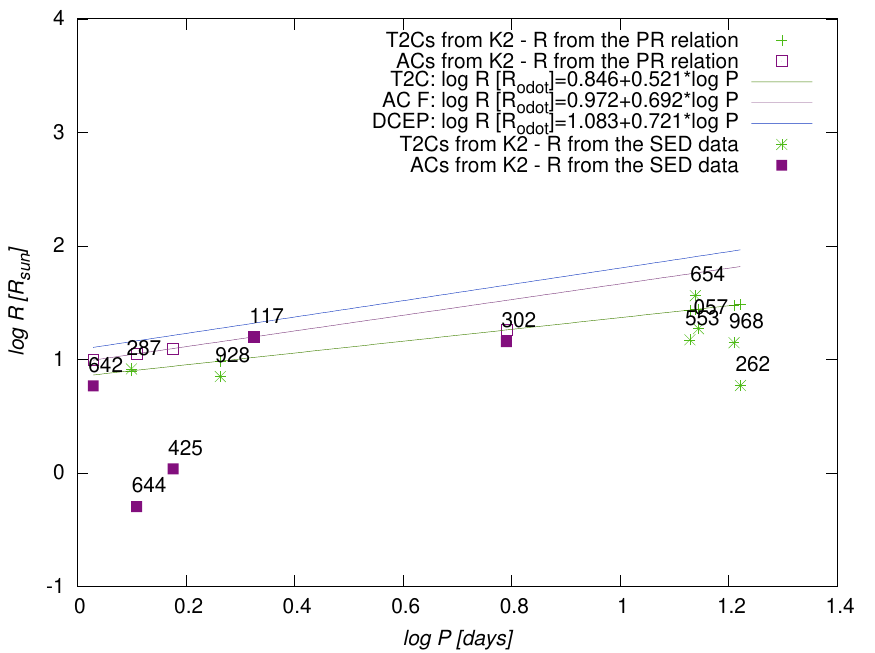}
    \caption{The period-radius relations for \textit{K2} T2Cs and ACs, with the used relations. The blue PR-relation is for the DCEP-F from \citet{2020A&A...635A..33G}: $\log{} R=(0.721\pm0.013) \log{} P+(1.083\pm0.012)$.}
    \label{fig:K2_R}
\end{figure}

The radius of the stars was calculated using the period-radius relations given in \citet{Groenewegen_2017b}, namely:

\begin{equation}
    \log R (R_{\sun})=0.846 \pm 0.006 + 0.521 \pm 0.006 \log P \mathrm{(d)},
	\label{eq:R_T2C}
\end{equation}

for the T2Cs, and

\begin{equation}
    \log R (R_{\sun})=0.972 \pm 0.005 + 0.692 \pm 0.034 \log P \mathrm{(d)},
	\label{eq:R_ACF}
\end{equation}

for the fundamental mode ACs.

We have also done a check of the radii by calculating it from the L and T$_{\textrm{eff}}$ which were obtained from the SED fits using the Stefan-Boltzmann law:

\begin{equation}
    L= 4 \pi R^2 \sigma T_{\textrm{eff}}^4,
    \label{eq:Stef-Bol}
\end{equation}

where L is the luminosity, R is the radius, $\sigma$ = 5.6704 10$^{-8}$ Wm$^{-2}$K$^{-4}$ is the Stefan-Boltzmann constant, and $T_{\textrm{eff}}$ is the effective temperature. All the resulting radii are given in Table~\ref{tab:radii}, and shown in Figure~\ref{fig:K2_R}. 

The PR-relation and the calculated radii are not model independent, one should be sceptical about using them as a parameter for classification. The $L$ calculated from the SED (given in Table~\ref{Tab:SED}) is significantly smaller than expected for an ACs, and thus this error is transferred to the radius calculation, as it can be seen in the cases of EPIC 246385425 and 246333644.

\begin{table}
	\centering
	\caption{The calculated radii for the stars in the \textit{K2} sample.}
	\label{tab:radii}
	\begin{tabular}{c c c}
	ID & log$R^{\mathrm{PR rel.}}$ & log$R^{\mathrm{SED}}$\\
	 & [R$_{\sun}$] & [R$_{\sun}$]\\
    \hline
    T2Cs & & \\
    \hline
    210622262 & 1.482 & 0.768 \\
    217235287 & 0.898 & 0.915 \\
    215881928 & 0.983 & 0.847 \\
    217987553 & 1.434 & 1.172 \\
    218642654 & 1.439 & 1.560 \\
    217693968 & 1.476 & 1.145 \\
    247445057 & 1.442 & 1.269 \\
    \hline
    ACs & & \\
    \hline
    202862302 & 1.258 & 1.155 \\
    218128117 & 1.198 & 1.193 \\
    246015642 & 0.993 & 0.765 \\
    246385425 & 1.094 & (0.033) \\
    246333644 & 1.048 & (-0.300) \\
 	\end{tabular}
\end{table}

%%%%%%%%%%%%%%%%%%%%%%%%%%%%%%%%%%%%%%%%%%%%%%%%%%%%%%%%%%%%%%%%

\section{The Period--Luminosity Relation}
\label{sec:PL_relation}

%%%%%%%%%%%%%%%%%%%%%%%%%%%%%%%%%%%%%%%%%%%%%%%%%%%%%%%%%%%%%%%%

The examined ACs and T2Cs form period-luminosity relations. In this section we test the limitation of the period-luminosity (magnitude) relations for the ACs and T2Cs (Subsection~\ref{subsec:PL_T2C_AC}) and the measured distances (Subsection~\ref{subsec:plx_vs_d}).

%%%%%%%%%%%%%%%%%%%%%%%%%%%%%%%%%%%%%%%%%%%%%%

\subsection{The period--magnitude relations of the Type II and anomalous Cepheids}
\label{subsec:PL_T2C_AC}

\begin{figure*}
    \centering
    \includegraphics[width=\textwidth]{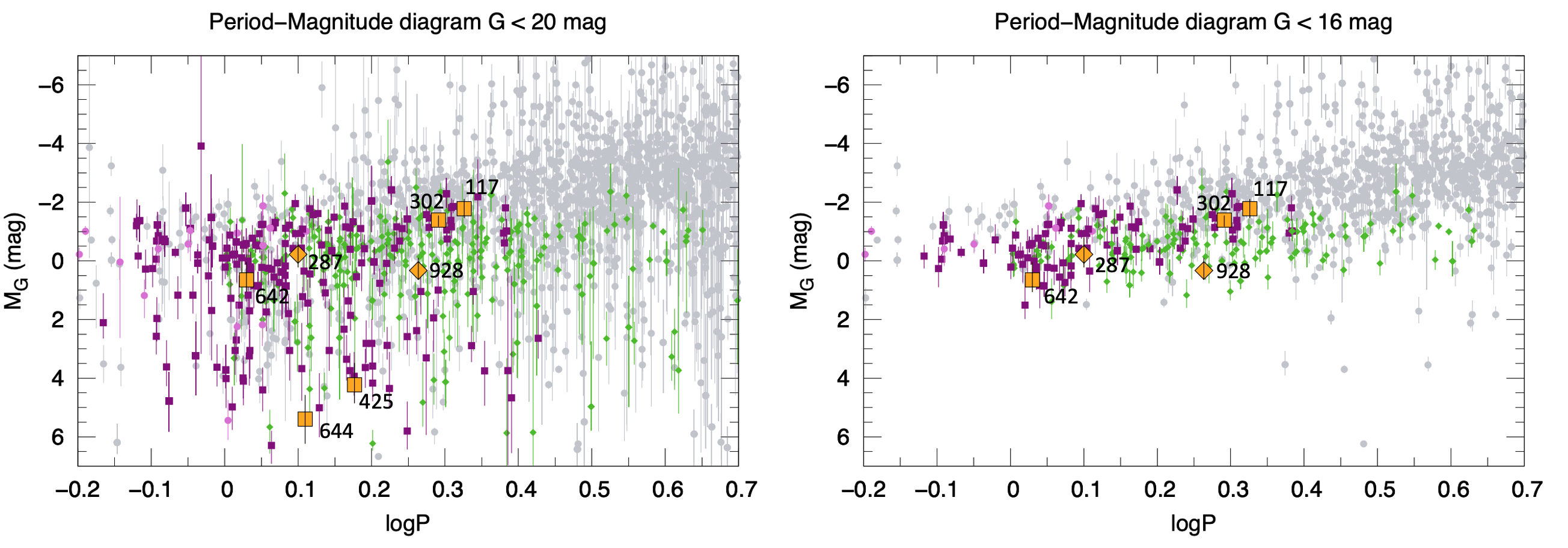}
    \caption{Period-magnitude diagrams for Cepheids with apparent brightness limit of 20 mag (left) and 16 mag (right) in Gaia $G$ band. Absolute brightness values were computed based on distances by \citep{Bailer-Jones_2018} considering interstellar dust from the \textit{Gaia} data base. The symbols are the same as in Figure~\ref{fig:CMD}.}
    \label{fig:pl_gaia}
\end{figure*}

The period-magnitude (luminosity) relation of pulsating stars is a method that could help us determine in an independent way to which subgroup these stars belong to: the ACs or the T2Cs.  We have taken the absolute \textit{Gaia} magnitudes using the EDR3 data for the T2Cs, ACs and DCEPs classified in the \citet{2019A&A...625A..14R} article and plotted the period--magnitude relation without a correction for the dust (see Figure~\ref{fig:pl_gaia}). In this general plot we can see that these three types of variables do not separate clearly, especially when the apparent brightness of the examined stars is fainter than 16 mag (see the left panel in Figure~\ref{fig:pl_gaia}). 

%%%%%%%%%%%%%%%%%%%%%%%%%%%%%%%%%%%%%%%%%%%%%%%%%%

\subsection{Parallax vs. distances}
\label{subsec:plx_vs_d}

\begin{figure*}
    \centering
    \includegraphics[width=1.0\textwidth]{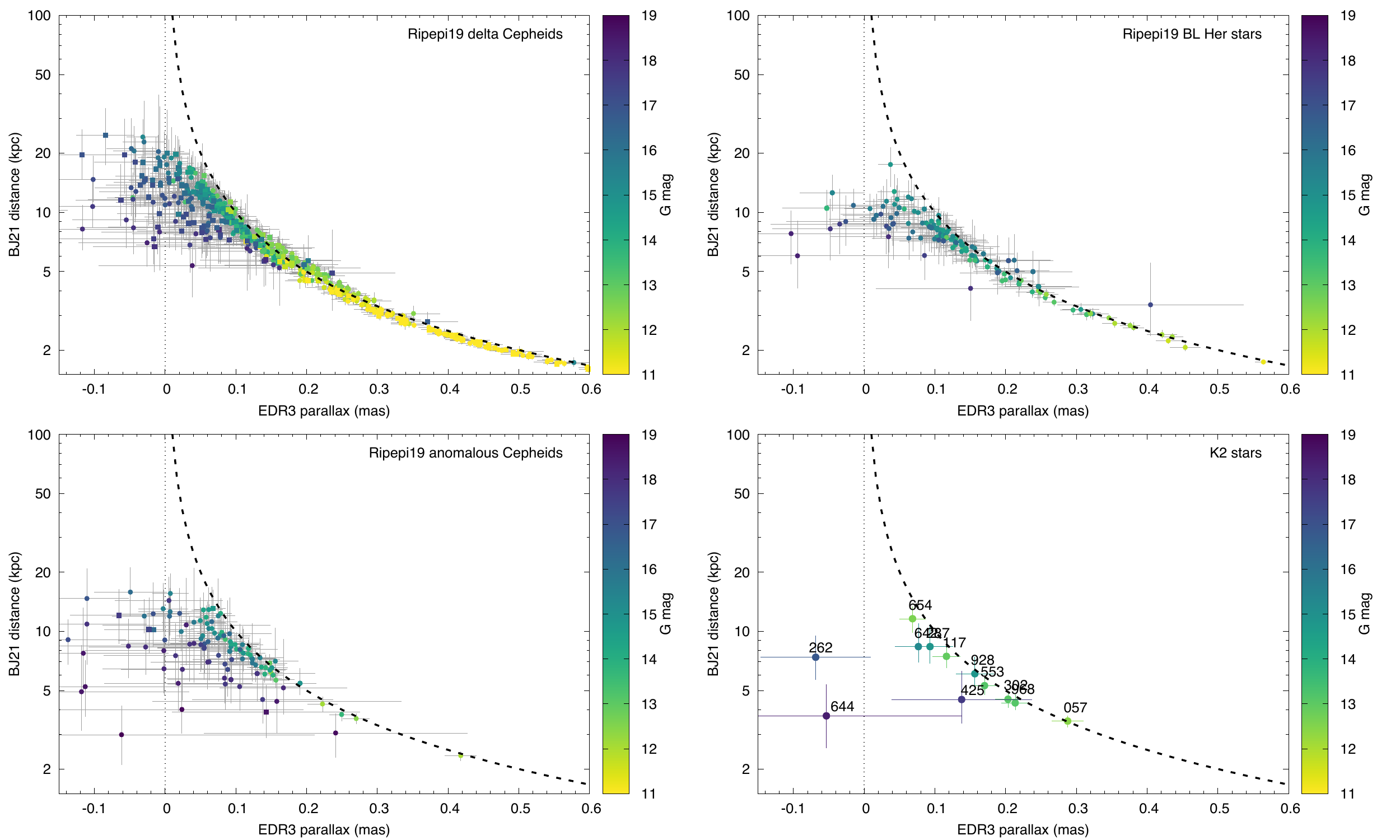}
    \caption{Gaia EDR3 parallaxes and geometric distances computed by \citet{BJ21} compared to each other. The black dashed line represents the simple $d=1/\pi$ inversion law. These plots clearly show that the calculated distances do not exceed 25 kpc, even for stars with very low parallaxes. For those stars the galactic model prior dominates the calculated distance, and it either underestimates the true distances for these stars or provides results that cannot be verified from the measured parallax.  }
    \label{fig:plx_vs_d}
\end{figure*}

Calculation of the absolute magnitudes requires the use of either a PL relation or some direct distance measurement, such as geometric parallaxes. The \textit{Gaia} mission is surveying the entire sky to collect astrometric and photometric data for nearly over a billion stars, and down to 20--21 magnitudes in the \textit{G} band \citep{GAIA_2016}. We collected the parallaxes from the \textbf{\textit{Gaia}} DR2 and EDR3 databases \citep{2018AA...616A...1G,2021A&A...649A...1G} and the geometric distances as derived from these data by \citet{Bailer-Jones_2018,BJ21}. However, some of our target stars are quite faint, which affects how useful their parallaxes might be. Since \textit{Gaia} is fundamentally a photometric mission, the accuracy of measuring stellar positions largely scales with brightness. Therefore the uncertainties in parallax are driven both by the brightness and distance of each target. Solutions fitted to position measurements with large fractional uncertainties could lead to highest-likelihood parallax values that are close to, or below zero. We can see this in the distribution of points from the Cepheid samples compiled by \citet{2019A&A...625A..14R} in Figure~\ref{fig:plx_vs_d}. 

As the \textit{Gaia} parallaxes are statistical quantities based on a set of measurements, we need to be careful in interpreting them as inverse distances. In order to derive distance estimates, \citet{Bailer-Jones_2018,BJ21} uses a detailed, three-dimensional model of the Galaxy as a prior in their calculations. This has the benefit of limiting distances to the expected extent of the Milky Way. However, when there is not enough information in the parallax data (due to near-zero or negative values, and/or high fractional uncertainties), this method will assign distances close to the densest, and thus most preferred part of the prior. We can see this if we compare the calculated geometric distances to the measured parallaxes in Figure~\ref{fig:plx_vs_d}. Here we applied the necessary parallax zero-point correction to the measurements \citep{Lindegren_2021}. The plots clearly show that that stars that are both faint and have small or negative parallaxes, concentrate between 5--20 kpc. Those distances are dominated by the prior, and thus cannot be reliably used for our purposes. 

If we now investigate the parallax and distance data for our target stars, we find that for the majority the two values agree with each other quite well. Only the three faintest targets are outliers. Out of those three, the anomalous Cepheid EPIC~246385425 appears to be in marginal agreement, while EPIC~210622262 and EPIC~246333644 have highly negative, and thus uninformative parallaxes. For those three stars absolute magnitudes cannot be reliably calculated from the \textit{Gaia} data.

%%%%%%%%%%%%%%%%%%%%%%%%%%%%%%%%%%%%%%%%%%%%%%%

\section{Conclusions}
\label{sec:conclusions}
%%%%%%%%%%%%%%%%%%%%%%%%%%%%%%%%%%%%%%%%%%%%%%%

Using the \textit{K2} data for the T2Cs and ACs we have performed our photometry with the autoEAP pipeline. Our final sample contained five ACs and seven T2Cs (two BLH and five WVir subtype stars). The classification was done using their Fourier parameters, which were compared to the $V$ band OGLE data of the ACs, T2Cs and DCEP stars. We have used Fourier analysis to examine the pulsational properties of the sample stars. The ACs and T2Cs showed only pulsation in the fundamental mode. Their O--C diagrams did not indicate any period change during the duration of the \textit{K2} Campaigns. None of the stars showed eclipses. In the O--C diagrams there was no evidence of binarity. The light curves did not show elliptical variability, which should be observed if the stars are in a close binary system. In the SED fits no IR-excess was observed, so there is no evidence of dust around the stars in the sample. The dust would be present in a close binary system exchanging material between the components. While the \textit{Gaia} RUWE parameter is not a definite show of binarity, it confirms that there are no observed binary stars among the stars in our sample, neither T2Cs nor ACs.

The BLH and AC stars in our sample are stable, as seen in Figure~\ref{fig:blh-ac}, where we present their light curves, phase folded light curves and O--C diagrams. We note that we do not see indication of Blazhko-like amplitude or phase modulation. O--C values were computed from the maxima of the pulsation cycles with a template fitting method. These values show a low scatter in the range of few minutes ($\sim10$ minutes at highest in EPIC 202862302). The pulsation amplitudes also vary within $\sim$5 mmag. At this low level we cannot decide weather these are instrumental or intrinsic, especially in the fainter stars where observational noise is significant (it can reach the 40 mmag at EPIC 246385425). Among WVir stars, EPIC 210622262 shows period-doubling, while the other WVir stars show cycle-to-cycle variation in their light curves.

We determined the luminosities, effective temperatures, and radii of the sample stars by modeling their SEDs with MoD. We used these parameters to place them on a HRD and analyze their evolution using the BaSTI models. In the case of the ACs the models with [Fe/H] of $-1.55$ and $-2.50$ and masses from 1.0 M$_{\odot}$ to 2.4 M$_{\odot}$ fit the region occupied by the ACs. For the T2Cs, we used three models, two with metal-poor metallicities, [Fe/H]=$-2.50$ and $-1.20$ and one with metallicity close to solar, [Fe/H]=$0.06$. The masses were from 0.45 to 0.8 M$_{\odot}$. The solar-like metallicity models fit the blue edge of the instability strip of the observed T2Cs. The CMD for ACs, T2Cs, and DCEPs from the \textit{Gaia} data does not clearly distinguish between these types of variables.

We show various ways that might be used to distinguish between T2Cs and ACs (and DCEPs, as well), making their classification unambiguous. Subsection~\ref{subsec:Fourier_analysis} shows the Fourier parameters and the light curves. We have used the available astrometric and photometric data from the \textit{Gaia} space telescope to construct the period--luminosity relation (Subsection~\ref{sec:PL_relation}) and the color--magnitude diagram (Subsection~\ref{subsec:CMD}). Using the results from the SED fitting we constructed the HRD for all of our stars in the sample. We have plotted the period-radius (PR-relation) for our stars.

For the T2Cs and ACs the most reliable classification method is based on their Fourier parameters (as we show in subsection~\ref{subsec:Fourier_analysis}) and their light curve shapes. Despite the advances in data precision even the \textit{Gaia} EDR3 data could not give a defining classification by plotting them on the color-magnitude diagram (see Subsection~\ref{subsec:CMD}) where the T2C, ACs, DCEP-F and DCEP-1O overlap. The same is true for the PL-relation, see Section~\ref{sec:PL_relation}.

We have note that stars EPIC 246385425 and 246333644 are outliers. The cause for this is connected to their observed brightness. This led us to examine the limitation of the distances estimated from the \textit{Gaia} space telescope. We confirm the findings from \citet{2021ApJS..253...11P} that for the pulsating stars that have a low apparent brightness the parallaxes are not good enough yet.

%%%%%%%%%%%%%%%%%%%%%%%%%%%%%%%%%%%%%%%%%%%%%%%%%%%%%%%%%%%%%%%%%%%%%%%%%%%%%%%%%%

\section*{Acknowledgements}
\label{sec:acknow}

This work has used \textit{K2} targets selected and proposed by the RR Lyrae and Cepheid Working Group of the \textit{Kepler} Asteroseismic Science Consortium (proposal numbers C4: GO4066, C7: GO7014, C10: GO10041, C12: GO12070, C13: GO13070). Funding for the \textit{Kepler} and \textit{K2} missions is provided by the NASA Science Mission directorate. 

Some of  the data presented in this paper were obtained from the Mikulski Archive for Space Telescopes (MAST). STScI is operated by the Association of Universities for Research in Astronomy, Inc., under NASA contract NAS5-26555. Support for MAST for non-HST data is provided by the NASA Office of Space Science via grant NNX13AC07G and by other grants and contracts.

This work has made use of data from the European Space Agency (ESA) mission {\it Gaia} (\url{https://www.cosmos.esa.int/gaia}), processed by the {\it Gaia} Data Processing and Analysis Consortium (DPAC, \url{https://www.cosmos.esa.int/web/gaia/dpac/consortium}). Funding for the DPAC has been provided by national institutions, in particular the institutions participating in the {\it Gaia} Multilateral Agreement.
This publication makes use of data products from the Two Micron All Sky Survey, which is a joint project of the University of Massachusetts and the Infrared Processing and Analysis Center/California Institute of Technology, funded by the National Aeronautics and Space Administration and the National Science Foundation.

The DENIS project has been partly funded by the SCIENCE and the HCM plans of the European Commission under grants CT920791 and CT940627. It is supported by INSU, MEN and CNRS in France, by the State of Baden-W\"urttemberg in Germany, by DGICYT in Spain, by CNR in Italy, by FFwFBWF in Austria, by FAPESP in Brazil, by OTKA grants F-4239 and F-013990 in Hungary, and by the ESO C\&EE grant A-04-046.

Jean Claude Renault from IAP was the Project manager. Observations were carried out thanks to the contribution of numerous students and young scientists from all involved institutes, under the supervision of  P. Fouqu\'e, survey astronomer resident in Chile.  

This publication makes use of data products from the Wide-field Infrared Survey Explorer, which is a joint project of the University of California, Los Angeles, and the Jet Propulsion Laboratory/California Institute of Technology, funded by the National Aeronautics and Space Administration. This work has made use of BaSTI web tools.

This project has been supported by the LP2012-31, LP2014-17, and LP2018-7 Lend\"ulet Programs of the Hungarian Academy of Sciences. MIJ acknowledges financial support from the Ministry of Education, Science and Technological Development of the Republic of Serbia through the contract number 451-03-68/2022-14/200002. This research was funded by DOMUS Grants of the Hungarian Academy of Sciences. Research was also supported by the D\'{e}lvid\'{e}k\'{e}rt Kiss Foundation, and the KKP-137523  `SeismoLab' \'Elvonal  grant  of  the  Hungarian  Research,  Development  and  Innovation  Office  (NKFIH). This work is supported by
the ERC via CoG-2016 RADIOSTAR (Grant Agreement 724560) and from work within the "ChETEC" (CA16117), and the "MW-Gaia" (CA18104) COST Actions funded by the COST program (European Cooperation in Science and Technology). 

\section*{Data availability}

The Fourier decomposition data underlying this article are available in the article and in its online supplementary material. The photometry data underlying this article will be shared on reasonable request by the corresponding author.

%%%%%%%%%%%%%%%%%%%%%%%%%%%%%%%%%%%%%%%%%%%%%%%%%%



%\bibliographystyle{mnras}

\bibliographystyle{mnras}
\bibliography{jurkovic_ref} 

%%%%%%%%%%%%%%%%%%%%%%%%%%%%%%%%%%%%%%%%%%%%%%%%%%

%%%%%%%%%%%%%%%%% APPENDICES %%%%%%%%%%%%%%%%%%%%%
%
\appendix
\section{Fourier parameters calculated from OGLE $V$ band measurements for LMC Cepheids}
\label{app:OGLE_V_Fourier}

The Fourier parameters for the pulsating stars from the OGLE $V$ band catalog were calculated using the same method as described in Section~\ref{sec:data}, so that our comparison on the Fourier parameter plots would be as calculated in a consistent manner. The calculated Fourier parameters are given in the Table~\ref{tab:ogle_fou} in a short format and are available fully online.

\begin{table*}
	\centering
	\caption{Fourier parameters calculated from OGLE $V$ band measurements for LMC Cepheids}
	\label{tab:ogle_fou}
	\begin{tabular}{lllllllllll}
Name	&	Type	&	Period (d)	&	R$_{21}$ 	&	
	$\sigma$R$_{21}$ 	&	$\phi_{21}$	&	 $\sigma\phi_{21}$ 	&	R$_{31}$ 	&	 $\sigma$R$_{31}$ 	&	$\phi_{31}$	&	 $\sigma\phi_{31}$ 	\\
		\hline
OGLE-LMC-CEP-0002	&	DCEP-F	&	3.118121	&	0.314	&	0.005	&	4.238	&	0.051	&	0.117	&	0.005	&	2.283	&	0.083	\\
OGLE-LMC-CEP-0012	&	DCEP-F	&	2.660176	&	0.460	&	0.003	&	3.984	&	0.024	&	0.290	&	0.002	&	2.032	&	0.036	\\
OGLE-LMC-CEP-0017	&	DCEP-F	&	3.677228	&	0.481	&	0.003	&	4.337	&	0.024	&	0.283	&	0.003	&	2.420	&	0.037	\\
OGLE-LMC-CEP-0023	&	DCEP-F	&	1.701824	&	0.478	&	0.006	&	3.780	&	0.046	&	0.163	&	0.005	&	1.820	&	0.072	\\
OGLE-LMC-CEP-0025	&	DCEP-F	&	3.733530	&	0.417	&	0.004	&	4.236	&	0.043	&	0.153	&	0.004	&	2.319	&	0.069	\\
OGLE-LMC-CEP-0026	&	DCEP-F	&	2.570668	&	0.455	&	0.003	&	4.057	&	0.016	&	0.234	&	0.002	&	1.961	&	0.025	\\
OGLE-LMC-CEP-0027	&	DCEP-F	&	3.522950	&	0.435	&	0.002	&	4.213	&	0.017	&	0.243	&	0.002	&	2.085	&	0.025	\\
OGLE-LMC-CEP-0028	&	DCEP-F	&	1.262950	&	0.468	&	0.004	&	3.770	&	0.035	&	0.299	&	0.005	&	1.552	&	0.051	\\
OGLE-LMC-CEP-0037	&	DCEP-F	&	3.066901	&	0.372	&	0.005	&	4.057	&	0.054	&	0.096	&	0.005	&	2.142	&	0.088	\\
OGLE-LMC-CEP-0039	&	DCEP-F	&	3.147752	&	0.403	&	0.004	&	4.246	&	0.041	&	0.203	&	0.004	&	2.261	&	0.065	\\
...	&	&	&	&	&	&	&	&	&	&	\\

	\hline

	\end{tabular}
\end{table*}

%%%%%%%%%%%%%%%%%%%%%%%%%%%%%%%%%%%%%%%%%%%%%%%%%%%%%%%%%%%%%%%%

\section{RV Tauri stars}
\label{sec:RVTs}

In the \textit{K2} proposals two RV Tauri stars were observed as shown in Table~\ref{tab:other_T2C}. The time span of observations in a given Campaign (less than 90 days) limits the scientific value of these long-period stars. These observations will be useful when additional data will be collected, so these data can contribute to the understanding of the pulsation of RV Tauri stars. The star EPIC 212146366 was observed in Campaigns 5 (two consecutive pulsational periods were covered) and 18. In case of EPIC 249577490 in Campaign 15 four pulsation periods were measured.

\begin{table*}
	\centering
	\caption{Type II Cepheids with long periods, which means that the cycle coverage was not enough for a detailed analysis of the pulsation. Reference $^1$: SIMBAD Astronomical Database (\url{http://simbad.u-strasbg.fr/simbad/})}
	\label{tab:other_T2C}
	\begin{tabular}{l l l l p{2.5cm} p{5.5cm}}
	C. & EPIC ID & RA [h m s] & DEC [$^\circ$ ' "] & Remark & Source\\
    \hline
    5, 18 & 212146366 & 08 26 10.154 & +22 57 17.81 & Redder T2C & \citet{2009AJ....137.4598S} (wrong period)\\
    \hline
    15 & 249577490 & 15 42 00.050 & -20 46 45.92 & RX Lib--RV Tau & SIMBAD$^1$\\
    \end{tabular}
\end{table*}

%%%%%%%%%%%%%%%%%%%%%%%%%%%%%%%%%%%%%%%%%%%%%%%%

%\section{Stars with variability different than pulsation of T2Cs and ACs}
\section{Non T2C and AC stars}
\label{sec:other}

The original TESS Asteroseismic Science Collaboration proposals for the \textit{K2} observing campaigns contained lists of pulsating stars that were collected from the literature.  Table~\ref{tab:other_stars} shows the list of stars that were among the observed stars, but after closer examination turned out not to be T2Cs or ACs. The light curves of these stars are quite diverse. In some cases we have determined that the variable in question is a DCEP (see 210426515, 211394018 and 247086981). There are two RRL type variable stars: 246395799 and 247520086. Some are changing their light curve due to being rotating spotted stars, ellipsoidal variables, eclipsing binaries. The star 246104577 is a triple star system with a debris disk. Some stars show flares in their light curve. Flares are not possible on T2Cs, so we have excluded them from the final list. Many of the observed stars are young stars (T Tauri type). There is a semi regular (SR), \textbf{an} irregular (L) and a RS CVn type variable as well. For the long period variable stars, typically 1.5-2.0 cycles of variation were observed in one campaign. 

\begin{table*}
	\centering
	\caption{Stars that were found to be classical Cepheids (DCEP), Cepheid variable stars (CEP), RR Lyrae variable stars (RRab and RRc), semi-regular variables (SR, SRD), BY Dra type variable stars (BY Dra), T Tauri type variables (T Tauri), young stellar objects (YSO), irregular variables (L), spotted stars, rotational variables (ROT), eclipsing binaries (EC), binary or triple stars, and others. These stars were observed in the examined Campaigns, and had light curves that could have resemble Type II Cepheids or anomalous Cepheids.}
	\label{tab:other_stars}
	\begin{tabular}{llllp{4.0cm}p{5.5cm}}
	C. & EPIC ID & RA [h m s] & DEC [$^\circ$ ' "] & Remark & Source\\
    \hline
	2 & 202571062 & 16 24 02.038 & -29 10 44.82 & Pre-main sequence EB & \citet{2016IBVS.6173....1V}\\
	2 & 204121113 &  16 09 58.630 & -23 34 55.91  & ROT & \citet{2018AJ....155..196R}\\
	2 & 204264833 & 16 23 07.782 & -23 00 59.58 & T Tauri & \citet{2016MNRAS.461..794P}\\
	2 & 204894575 &  16 02 53.963 & -20 22 48.00 & T Tauri & \citet{2016MNRAS.461..794P}\\
	2 & 205023358 & 15 55 06.701 & -19 46 31.33 & ROT or BY Dra & \citet{2012AcA....62...67K} or \citet{2003AstL...29..468S}\\
	\hline
	3 & 205964079 & 22 24 45.925 & -15 54 48.32 & Spotted flare star & This article\\
	\hline
	4 & 210990639 & 04 11 08.549 & +22 49 31.31 & Binary (eclipse seen) & This article\\
	4 & 210426515 & 03 49 17.915 & +14 03 14.59 & ROT: & ASAS-SN \citep{2017PASP..129j4502K}\\
	4 & 210525462 & 04 05 26.387 & +15 49 49.88 & L & ASAS-SN \citep{2018MNRAS.477.3145J}\\
	4 & 210708282 & 03 28 08.580 & +18 28 54.71 & ROT or EC/CW-FO/ESD & \citet{2010MNRAS.408..475H} or \citet{2002AcA....52..397P}\\
	4 & 210785321 & 03 28 13.447 & +19 38 44.96 & L & ASAS-SN \citep{2019MNRAS.486.1907J}\\
	4 & 210800332 & 04 02 59.906 & +19 52 20.76 & MISC/CW & \citet{2002AcA....52..397P}\\
	\hline
	5 & 211394018 & 08 35 22.466 & +11 32 57.93 & DCEP & \citet{2011AJ....141...53S}\\
	5, 18 & 211584699 & 08 12 49.000 & +14 19 52.89 & L & ASAS-SN \citet{2019MNRAS.486.1907J}\\
	5 & 211759736 & 08 15 12.967 & +16 44 41.47 & Spotted giant star in an EC & \citet{2018AA...620A.189O}\\
	\hline
	7 & 214047277 & 18 56 17.534 & -27 31 25.62 & V4061 Sgr--SRD & ASAS-SN \citep{2019MNRAS.486.1907J}\\
	7 & 217257168 & 18 57 51.984 & -20 53 37.05 & SR &  \citet{2017ARep...61...80S}\\
	7 & 218177215 & 18 56 25.238 & -19 16 22.87 & Dwarf nova & This article\\
	\hline
	10 & 201419619 & 12 07 56.712 & -00 39 37.62 & Binary star & This article\\	
	10 & 228811074 & 12 28 12.493 & -06 19 00.53 & ROT with flares & This article\\
	\hline
	12, 19 & 246104577 & 22 59 34.998 & -07 02 22.87 & Triple system with debris disk & \citet{2013MNRAS.434.1117D}\\
	12 & 246263296 & 23 36 06.012 & -03 47 47.82 & ROT with spots and flares & This article,\\
	 & & & & & ROT--ASAS-SN \citep{10.1093/mnras/stz2711}\\
	12 & 246395799 & 23 13 55.646 & -01 07 25.08 & RRc & This article\\
	\hline
	13 & 210670948 & 04 32 09.287 & +17 57 22.69 & Weak T Tauri & \citet{1999AJ....118.1043H}\\
	13 & 210698281 & 04 32 14.568 & +18 20 14.84 & Weak T Tauri & \citet{1999AJ....118.1043H}\\
	13 & 246736776 & 05 01 29.240 & +15 01 26.35 & ROT & ASAS-SN \citep{2019MNRAS.486.1907J}\\
	13 & 246773415 & 05 00 49.288 & +15 27 00.72 & YSO & \citet{2016AA...594A..63G}\\
	13 & 246782263 & 04 32 56.252 & +15 32 53.20 & Spotted, flare star with eclipses & This article\\
	13 & 247086981&	04 37 14.779 & +18 32 34.92 & SZ Tau--DCEP & \citet{2015MNRAS.446.4008E}\\ 
	13 & 247454835 & 04 31 16.860 & +21 50 25.26 & T Tau type & \citet{2019MNRAS.483.1642H}\\
	13 & 247520086 & 04 58 08.497 & +22 20 59.86 & RRab & \citet{2013ApJ...763...32D}\\
	13 & 247671949 & 04 57 47.113 & +23 30 26.25 & ROT: & ASAS-SN \citep{2019MNRAS.486.1907J}\\
	\hline
	15 & 249104247 & 15 22 16.298 & -26 52 25.54 & RS CVn (X-ray source) & PX Lib \citet{2008PZ.....28....9B}\\
	15 & 249483655 & 15 26 02.036 & -21 56 36.24 & Eclipsing ellipsoidal binary & This article\\
	 & & & & & ROT--ASAS-SN \citep{2019MNRAS.486.1907J}\\
	15 & 249926841 & 15 30 48.035 & -16 19 23.04 & Spotted, flare star & This article,\\
	 & & & & & ROT--ASAS-SN \citep{2019MNRAS.486.1907J}\\
	15 & 250178428 & 15 29 43.29 & -13 10 52.00 & YSO & ASAS-SN \citep{2019MNRAS.486.1907J}\\
	\hline
	19 & 251737903 & 22 55 18.04 &	+01 11 53.62 & Galaxy & \citet{2010MNRAS.405..783M}\\ 
	\end{tabular}
\end{table*}

%%%%%%%%%%%%%%%%%%%%%%%%%%%%%%%%%%%%%%%%%%%%%%%


%%%%%%%%%%%%%%%%%%%%%%%%%%%%%%%%%%%%%%%%%%%%%%%%%%

% Don't change these lines
\bsp	% typesetting comment
\label{lastpage}
\end{document}